\documentclass[journal]{IEEEtran}
\IEEEoverridecommandlockouts

\usepackage{amsmath}
\usepackage[linesnumbered, ruled]{algorithm2e}
\usepackage{amsfonts}
\usepackage{amsthm}
\usepackage{algorithmicx}
\usepackage{array}
\usepackage[caption=false,font=normalsize,labelfont=sf,textfont=sf]{subfig}
\usepackage{textcomp}
\usepackage{stfloats}
\usepackage{url}
\usepackage{verbatim}
\usepackage{graphicx}  
\usepackage{float}  
\usepackage{subfloat}  

\usepackage{cite}
\usepackage{multicol}

\usepackage{bm}
\usepackage{booktabs}
\usepackage{amssymb}
\usepackage{comment}
\usepackage[colorlinks=true, allcolors=blue]{hyperref}
\usepackage{mathrsfs}
\usepackage{tikz}
\usepackage{pifont}

\begin{document}

\title{Power Optimization in Multi-IRS Aided Delay-Constrained IoVT Systems
	{\footnotesize \textsuperscript{}}

}

\author{Baolin Chong, Hancheng Lu,~\IEEEmembership{Senior Member,~IEEE}, Langtian Qin, Chenwu Zhang, Jiasen Li and Chang Wen Chen,~\IEEEmembership{Fellow,~IEEE}
\IEEEcompsocitemizethanks{\IEEEcompsocthanksitem
This work was supported  in part by the Hong Kong Research Grants Council under Grant GRF-15213322 and in part by the National Science Foundation of China under Grant U21A20452 and Grant U19B2044.

Baolin Chong, Langtian Qin, Chenwu Zhang, and, Jiasen Li  are with the CAS Key Laboratory of Wireless-Optical Communications, University of Science and Technology of China, Hefei 230027, China. (e-mail: chongbaolin@mail.ustc.edu.cn; qlt315@mail@mail.edu.cn; cwzhang@ustc.edu.cn; JasonL@mail.ustc.edu.cn).

Hancheng Lu is with the CAS Key Laboratory of Wireless-Optical Communications, University of Science and Technology of China, Hefei 230027, China,  and also with the Hefei Comprehensive National Science Center, Institute of Artificial Intelligence, Hefei 230027, China. (e-mail: hclu@ustc.edu.cn).

Chang Wen Chen is with the Department of Computing, The Hong Kong Polytechnic University, Hong Kong. (e-mail: changwen.chen@polyu.edu.hk). }
}

\maketitle{}

\begin{abstract}
With the advancement of video sensors in the Internet of Things, Internet of Video Things (IoVT) systems, capable of delivering abundant and diverse information, have been increasingly deployed for various applications.
However, the extensive transmission of video data in IoVT poses challenges in terms of delay and power consumption.
Intelligent reconfigurable surface (IRS), as an emerging technology, can enhance communication quality and consequently improve system performance by reconfiguring wireless propagation environments.
Inspired by this, we propose a multi-IRS aided IoVT system that leverages IRS to enhance communication quality, thereby reducing power consumption while satisfying delay requirements.
To fully leverage the benefits of IRS, we jointly optimize power control for IoVT devices and passive beamforming for IRS to minimize long-term total power consumption under delay constraints.
To solve this problem, we first utilize Lyapunov optimization to decouple the long-term optimization problem into each time slot. Subsequently, an alternating optimization algorithm employing optimal solution-seeking and fractional programming is proposed to effectively solve the optimization problems at each time slot.
Simulation results demonstrate that the proposed algorithm significantly outperforms benchmark algorithms in terms of long-term total power consumption.
Moreover, a trade-off between the number of IRS elements and system performance is also proved.
\end{abstract}


\section{Introduction}
\IEEEPARstart{W}{ith} the proliferation of video sensors, such as three-dimensional (3D) cameras and camera arrays, which can provide richer and more diverse video data in the Internet of Things (IoT), a new sub-field of IoT, called the Internet of Video Things (IoVT) \cite{ChenInternet2020}, has had a significant impact on various applications, such as public safety \cite{ChenDynamic2016}, traffic analysis \cite{JiCrowd2020}, and industrial automation \cite{GuiRobust2021}.
Compared to conventional IoT systems, IoVT systems present novel challenges.
Firstly, unlike structured data, ensuring the real-time transmission of video content mandates the imperative to guarantee low latency in the uploading of data from IoVT devices \cite{KhalekDelay2015}. Furthermore, the transmission of massive volumes of video data also imposes heightened requirements on energy consumption.

Numerous research efforts have been made to enhance service quality in the IoVT system. Mobile edge computing is employed to offload the substantial video data generated by IoVT devices to servers for processing, ensuring minimal latency \cite{GuoNOMA2021, ChenJoint2022, YangJoint2022}.
Furthermore, non-orthogonal multiple access (NOMA) technology is employed for uplink data transmission in IoVT systems \cite{GuoNOMA2021}, \cite{YangJoint2022}, \cite{MaQoS2021}, enhancing uplink throughput through superimposed coding and successive interference cancellation, thus reducing latency in video data transmission.
The aforementioned endeavor has been validated to significantly enhance the performance of IoVT systems under the influence of channel attenuation caused by adverse radio propagation conditions.

The attenuation of signals by wireless channels profoundly affects communication quality, thereby limiting the capabilities of IoVT systems. Intelligent reflecting surfaces (IRS), as an emerging technology, have the potential to mitigate the effects of channel attenuation by reconfiguring the wireless propagation environment \cite{BasarWireless2019}, \cite{PanReconfigurable2021}.
IRS consists of a large number of passive reflecting elements, each of which can independently cause amplitude and phase changes in the electromagnetic wave. By carefully designing the amplitude and phase of all reflecting elements, the reflected electromagnetic wave can be coherently superimposed with the direct signal from the IoVT device to achieve passive beamforming (PBF), thereby enhancing the strength of the signal.
On the one hand, signal enhancement elevates the rate of video data transmission, ensuring low-latency propagation of video data. On the other hand, this signal enhancement correspondingly diminishes the demand for transmit power of IoVT devices, thereby achieving energy efficiency within the IoVT system.
In existing research, IRS has been widely applied to enhance useful signal power, such as in severely blocked line-of-sight (LOS) links \cite{ZhaoJoint2022}, \cite{ZhaoJoint2020}, improving IRS-assisted UAV networks \cite{PangWhen2021}, \cite{PangIRS2022}, enhancing channel capacity \cite{ZhangA2021, ZhaoReconfigurable2021, LuJoint2022}, and reducing transmit power \cite{WuBeamforming2020}.

However, enabling the utilization of multi-IRS to support delay-constrained IoVT systems is challenging where both the power control at the IoVT device and PBF at IRS need to be carefully designed to achieve desired system performance. On the one hand, the introduction of the IRS introduces novel optimization variables, where the channels designed by the IRS are cascaded and influenced by the IRS reflection matrix, resulting in the coupling of the IRS reflection matrix with power control, which needs to be jointly designed. On the other hand, the power control and PBF across different time slots are coupled, jointly impacting system performance and necessitating simultaneous consideration to meet delay requirements.
In this paper, we propose a multi-IRS aided IoVT system, where multiple IRSs assist in transmitting video data for all IoVT devices under delay constraints. To achieve low-power communication,  we attempt to minimize long-term total power consumption by solving the joint optimization problem on power control and PBF.
Although many works have already studied the joint power control and IRS PBF problems in different scenarios, such as multiple-input single-output system \cite{HuangReconfigurable2019}, NOMA \cite{XieJoint2021}, simultaneous wireless information and power transfer networks \cite{LiJoint2022}, and user-centric network \cite{ LuJoint2022}.
However, within a delay-constrained IoVT system, it is crucial to consider the long-term impact of the proposed solution to ensure the satisfaction of delay requirements. It is challenging to jointly optimize the power of IoVT devices and phases of multi-IRS across interconnected time slots.
Our main contributions can be summarized in the following three aspects:
\begin{itemize}
  \item We propose a multi-IRS aided IoVT system where multiple IRSs are deployed to assist IoVT devices for uplink video data transmission. Then, by jointly considering the transmit power control and IRS reflection matrix, we formulate a delay-constrained long-term total power consumption minimization problem.
  \item To ensure delay without relying on future information and simultaneously achieve the long-term goal of minimizing total power consumption, we employ Lyapunov optimization to decouple the original long-term optimization problem into the optimization problem at each time slot. Since the optimization problem is non-convex at each time slot, we propose a two-step alternating optimization algorithm to address this challenge. Specifically, the original problem is firstly decomposed into two sub-problems, involving uplink power control and IRS reflection matrix design. Subsequently, these two problems are iteratively solved by calculating the optimal scheme and fractional programming (FP) method in an alternating manner until the objective function converges to a stable value.
  \item Extensive simulations are performed to validate that the utilization of IRS in delay-constrained IoVT systems can effectively improve network performance when compared to existing algorithms. The results show that the proposed algorithm outperforms benchmark algorithms and achieves significant gains in terms of total power consumption. Moreover, we find that there is a trade-off between the number of IRS elements and system performance. By selecting the most appropriate quantity of IRS elements in practical implementations, it becomes possible to achieve the lowest total power consumption while satisfying the delay requirement.
\end{itemize}

The rest of the paper is organized as follows. Section II introduces the system model and problem formulation. Then, Lyapunov optimization is used to decouple the long-term total power consumption minimization problem in Section III. In Section IV, we propose an alternating optimization algorithm for solving the problem at each time slot. Section V discusses details about the channel model, simulation setup, and numerical results. Finally, the conclusion and future work are drawn in Section VI.

\emph{Notations:} In this paper, vectors, and matrices are denoted by lowercase and uppercase bold letters, respectively.
For a general matrix $\mathbf{A}$, $\mathbf{A}^H$ and $\mathbf{A}^{-1}$represent the Hermitian and inverse of $\mathbf{A}$, respectively.
$\left | \cdot \right | $ and $\left ( \cdot \right )^* $ denote the modulus and the conjugate of a complex number, respectively. $\mathbb{C}^{x\times y}$ represent the space of $x\times y$ complex number matrices.
$\mathbb{E}\left \{ x \right \}$ denote the expected value of $x$ and $j \triangleq \sqrt{-1}$ presents the imaginary unit.

\section{System Model and Problem Formulation}\label{Section2}
\begin{figure}[ht]
	\centering
	\includegraphics[scale=0.35]{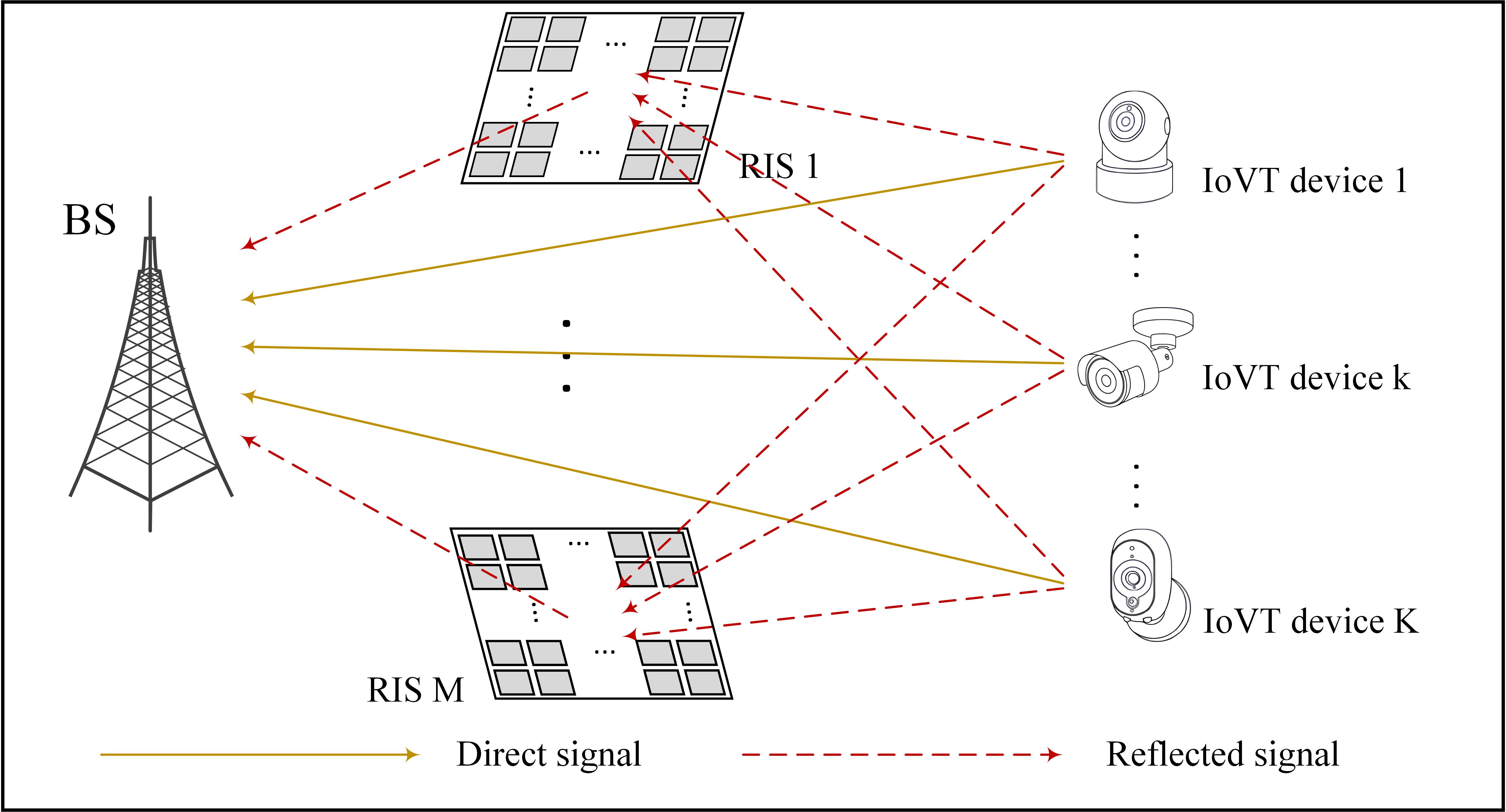}
	\caption{A multi-IRS assisted IOVT system.}
	\label{fig:model}
\end{figure}

In this section, we first describe the system model as shown in Fig. \ref{fig:model}, where multiple IRSs are deployed to collaboratively serve multiple IoVT devices. Subsequently, we aim to minimize the long-term total power consumption during the uplink video data transmission phase while satisfying the delay constraints. The problem formulation encompasses the joint optimization of both power control and the design of the IRS reflection matrix.

\subsection{System Model}
We consider an uplink multi-IRS aided IoVT system, comprising a base station (BS), $M$ IRS, and $K$ IoVT devices.
In this system, strategically deploying IRS on building surfaces allows for the reconstruction of the wireless propagation environment between IoVT devices and the BS, thereby enhancing communication quality.
The adopted framework follows a time-slotted model with a uniform slot duration denoted as $\tau$, encompassing a total of $T$ time slots.
For convenience, we define the index sets $\mathcal{M} \triangleq \{1, 2, \cdots, M\}$, $\mathcal{K} \triangleq \{1, 2, \cdots, K\}$, and $\mathcal{T} \triangleq \{1, 2, \cdots, T\}$ to represent the sets of IRSs, IoVT devices, and time slots, respectively.

Each IRS within the proposed system is equipped with $N$ reflecting elements.
Mathematically, the signal reflected by the IRS is characterized by the multiplication of the incident signal with a matrix of complex reflection coefficients.
Let $\pmb{\Phi}_m^t = \text{diag}\left \{ \beta_{m,1}^te^{j\phi_{m,1}^t},\cdots ,\beta_{m,N}^te^{j\phi_{m,N}^t} \right \}\in \mathbb{C}^{N\times N} $ represent the reflection matrix of the $m$-th IRS in the $t$-th time slot, where $\beta_{m,n}^t$ and $\phi_{m,n}^t$ denote the amplitude coefficient and the phase shift of the $n$-th element of the $m$-th IRS in the $t$-th time slot, $\forall m \in \mathcal{M}$, $\forall t \in \mathcal{T}$. To maximize the reflection signal, a universal setting of $\beta_{m,n}^t = 1$ is adopted for all IRSs, reflecting elements, and time slots during the uplink transmission phase \cite{WuIntelligent2019}. Furthermore, the utilization of discrete phase shifts is contemplated, primarily due to their greater cost-effectiveness in comparison to continuous phase adjustments. Each reflecting element of the IRS has $2^b$ levels, the set of discrete phase-shift values is $\left \{ 0, \Delta\theta,\cdots,\left ( 2^b-1 \right )\Delta\theta  \right \} $, where $\Delta\theta = 2\pi/2^b $.

During the uplink transmission of video data, we assume that each IRS reflects signals from any IoVT devices within the system. The intensity of the reflected signal is influenced by various factors, including the distances between entities \cite{ZhangA2021}.
Considering that the cumulative path loss of IRS-reflected links arises from the interplay of distances, signals undergoing reflection by two or more IRSs are omitted from our consideration.
Accordingly, the channel between BS and each IoVT device consists of the direct BS-device channel and $M$ reflected BS-IRS-device channels. Then, the channel between the BS and the $k$-th IoVT device in the $t$-th time slot is given by
\begin{equation}\label{overall channel}
  \begin{aligned}
    h_k^t &= h_{d,k}^t + \sum_{m\in \mathcal{M}} \left ( \mathbf{f}_{m}^t \right )^H \pmb{\Phi}_m^t\mathbf{g}_{m,k}^t\\
          &= h_{d,k}^t + \left ( \mathbf{f}^t \right )^H \pmb{\Phi}^t \mathbf{g}_{k}^t,
  \end{aligned}
\end{equation}
where $h_{d,k}^t $ denotes the direct channel between the BS and the $k$-th IoVT device in the $t$-th time slot, $\mathbf{f}_{m}^t \in \mathbb{C}^{N\times 1}$ denotes the channel from the BS to the $m$-th IRS in the $t$-th time slot, and $\mathbf{g}_{m,k}^t \in \mathbb{C}^{N\times 1}$ represents the channel that from the $m$-th IRS to the $k$-th IoVT devices in the $t$-th time slot.
To enhance the visual clarity of the channel composition, we additionally introduce definitions $ \mathbf{f}^t \in \mathbb{C}^{MN\times 1}$ and $\mathbf{g}_{k}^t \in \mathbb{C}^{MN\times 1}$ as the equivalent channel from the BS to all IRSs and the channel from all IRSs to the $k$-th IoVT devices in the $t$-th time slot, respectively.
Let $\pmb{\Phi}^t = \text{diag}\left \{ \pmb{\Phi}_1^t,\cdots,\pmb{\Phi}_M^t \right \}  \in \mathbb{C}^{MN\times MN}$ represents the equivalent reflection matrix of all IRS in the $t$-th time slot, $\mathbf{v}^t = \left [e^{j\phi_{1,1}^t},\cdots ,e^{j\phi_{1,N}^t},\cdots,e^{j\phi_{L,1}^t},\cdots,e^{j\phi_{L,N}^t} \right ]^T\in \mathbb{C}^{MN\times 1}$ denotes the reflection phase vector of all IRS without the consideration of reflection amplitude in the $t$-th time slot. The channel $\mathbf{h}_{c,k}^t = \text{diag}\left \{ \mathbf{g}_{k}^t \right \}\mathbf{f}^t  \in \mathbb{C}^{MN\times 1}$ denotes the cascaded channel from the BS to the $k$-th IoVT device via all IRSs in the $t$-th time slot. Accordingly, $h_k^t$ can be equivalently expressed as
\begin{equation}\label{represent of overall channel}
  \begin{aligned}
    h_k^t = h_{d,k}^t + \left ( \mathbf{v}^t \right )^H \mathbf{h}_{c,k}^t.
  \end{aligned}
\end{equation}

During uplink video data transmission, the orthogonal frequency division multiplexing technique is adopted \cite{RenIntelligent2022, HuangCombining2023}, where all IoVT devices communicate simultaneously with the BS on different orthogonal sub-carriers of equal size and the bandwidth of each sub-carrier is $B$. Therefore, the achievable data rate from the $k$-th IoVT device to the BS in the $t$-th time slot is expressed as
\begin{equation}\label{acvhievable rate}
  \begin{aligned}
    R_k^t = B \log_2\left ( 1+\frac{p_k^t \left | h_k^t \right |^2}{\sigma^2}  \right ),
  \end{aligned}
\end{equation}
where $p_k^t$ denotes the transmit power of the $k$-th IoVT devices in the $t$-th time slot, and $\sigma^2$ represents the variance of noise.

A data buffer is employed to store the uplink video data at each IoVT device. We assume that video data, with magnitudes denoted by $\mathbf{A}^t = [ A_1^t, A_2^t, \cdots, A_K^t ] ^T $ (in bits), arrives at the queues of IoVT devices during the $t$-th time slot.
Besides, we assume that $A_k^t \in \left [ A_{k, min}^t,A_{k,max}^t \right ] $, $\forall k \in \mathcal{K}$, follows an independent and identically uniformed distribution across distinct time slots, encompassing non-negative integer values. Consequently, we define $\mathbf{Q}^t = [ Q_1^t, Q_2^t, \cdots, Q_K^t ]$ to represent the queue lengths of all IoVT devices at the beginning of the $t$-th time slot. Therefore, the queue backlog of the $k$-th IoVT device is updated as follows:
\begin{equation}\label{queue update}
  \begin{aligned}
    Q_k^{t+1} = \max \left \{ Q_k^t +A_k^t - R_k^t \tau,0  \right \}.
  \end{aligned}
\end{equation}
Note that the queue of each IoVT device is empty at the beginning, i.e., $Q_k^1 = 0$, $\forall k \in \mathcal{K}$.

The total power consumption of the proposed multi-IRS aided IoVT system consists of the transmit power of IoVT devices and the static power of the hardware consumption consumed by IRSs. Other aspects of power consumption are considered fixed in this paper and thus excluded from consideration. In the $t$-th time slot, the total power consumption $P^t$ is composed of the transmit power $p_k^t$ of each IoVT device and the power consumption of each IRS element $P_{I,n}^t$, which can be expressed as
\begin{equation}\label{total powe consumption of system}
  \begin{aligned}
    P^t= \sum_{k \in \mathcal{K}} p_k^t + MNP_{I,n}^t.
  \end{aligned}
\end{equation}

\subsection{Problem Formulation}

According to Little's Law, the average queuing delay within an IoVT device's buffer is directly linked to the ratio of the average queue length to the corresponding average data arrival rate \cite{LiuDynamic2019, ross2014introduction}. Therefore, the average queuing delay of the $k$-th IoVT device is represented as
\begin{equation}\label{average queue delay}
  \begin{aligned}
    d_k^{av} = \lim_{T \to +\infty}\frac{1}{T}  \sum_{t=1}^{T}d_k^t = \lim_{T \to +\infty}\frac{1}{T} \sum_{t=1}^{T}\frac{Q_k^t}{\tilde{A}_k^{t-1}},
  \end{aligned}
\end{equation}
where $d_k^t = {Q_k^t}/{\tilde{A}_k^{t-1}}$, and ${\tilde{A}_k^{t}} = \frac{1}{t}  {\textstyle \sum_{i=1}^{t}} A_k^i$ represents the time-averaged data arrival rate in the $t$-th time slot.

In the proposed multi-IRS aided IoVT system, we formulate the problem of long-term total power consumption minimization considering the jointly optimizing the IRS reflection matrix (i.e., $\mathbf{v} = \left \{ \mathbf{v}^1,\mathbf{v}^2,\cdots,\mathbf{v}^t,\cdots \right \}$), and the transmit
power (i.e., $\mathbf{p} =  \left \{ \mathbf{p}^1,\mathbf{p}^2,\cdots,\mathbf{p}^t,\cdots \right \}$ across sequential time slots while satisfying delay constrained, where $\mathbf{p}^t = \left [ p_1^t,p_2^t,\cdots,p_K^t \right ]^T$ denotes the transmit power of IoVT devices in the $t$-th time slot. Mathematically, the optimization problem can be formulated as
\begin{subequations}\label{orignal optimization problem}
\begin{align}
\min_{\mathbf{v}, \mathbf{p}}\ \lim_{T \to \infty}\frac{1}{T} \sum_{t\in \mathcal{T}} \left ( \sum_{k\in\mathcal{K}}  p_k^t  + MNP_{I,n}^t  \right )  \ \ \label{objection function 1}
\end{align}
\begin{alignat}{2}
\text{s.t.}\
             &\phi_{m,n}^t \in \left \{ 0, \Delta\theta,\cdots,\left ( 2^b-1 \right )\Delta\theta  \right \}, \nonumber\\
             &\quad \forall m\in \mathcal{M},n\in \mathcal{N} ,t\in\mathcal{T} ,   \label{orignal constrited 1} \\
             &0 \le p_k^t \le p_{k,max},\quad  \forall k \in \mathcal{K},t\in\mathcal{T} ,   \label{orignal constrited 2} \\
             & \lim_{T \to +\infty} \frac{1}{T} \sum_{t \in \mathcal{T}} d_k^t \le d_k^{th},\quad \forall k \in \mathcal{K},  \label{orignal constrited 4}
\end{alignat}
\end{subequations}
where $p_{k,max}$ denotes the maximal transmit power of the $k$-th IoVT device, $d_k^{th}$ is the delay threshold of $k$-th IoVT device, constraint (\ref{orignal constrited 1}) ensures that each IRS reflecting element only provides a discrete phase shift, constraint (\ref{orignal constrited 2}) accounts for the fact that transit power of all IoVT device is kept below the maximum power, and constraint (\ref{orignal constrited 4}) ensures that the average queuing delay of each IoVT device is below the threshold.

\section{Problem Decoupling}
At each time slot, we make decisions without prior knowledge of future channel conditions and data arrival patterns to minimize long-term power consumption while simultaneously ensuring that average delay requirements are met. To address this challenge, we employ Lyapunov optimization, which allows us to decouple the long-term optimization problem (\ref{orignal optimization problem}) into a series of independent optimization problems across different time slots \cite{NeelyEnergy2006, LiuOnline2018}.
The specific details of using Lyapunov optimization to solve the problem (\ref{orignal optimization problem}) are introduced in the following.

We first use the Lyapunov drift-plus-penalty method to transform the average delay constraint (\ref{orignal constrited 4}) into a queue stability constraint.
Specifically, denote $\left \{ D_k^t \right \}_{k \in \mathcal{K} }$ as the virtual queues for all IoVT devices in the $t$-th time slot, and the virtual queues are updated at each time slot as
\begin{equation}\label{virtual queue update}
  \begin{aligned}
    D_k^{t+1} = \max \left \{ D_k^{t} - d_k^{th} + d_k^{t+1},0 \right \},\ \forall k\in\mathcal{K}.
  \end{aligned}
\end{equation}
Note that when the long-term average length of the virtual queues is finite, those virtual queues can be referred to as strongly stable, i.e.,
\begin{equation}\label{stable condition}
  \begin{aligned}
    \lim_{T \to \infty} \frac{1}{T} \sum_{t=0}^{T-1} \mathbb{E}\left \{ D_k^t \right \}  < \infty,\ \forall k \in \mathcal{K}.
  \end{aligned}
\end{equation}
To ensure that the average delay constraints for all IoVT devices are met, it is necessary for their corresponding virtual queues to be strongly stable.

Subsequently, we ensure the stability of the virtual queue lengths to guarantee the fulfillment of average delay constraints.
Define the quadratic Lyapunov function in the $t$-th time slot as
\begin{equation}
  \begin{aligned}
    L\left ( \mathbf{D}^t \right ) = \frac{1}{2}   \sum_{k \in \mathcal{K} }  \left ( D_k^t \right )^2.
  \end{aligned}
\end{equation}
where $\mathbf{D}^t = \left [ D_1^t,D_2^t,\cdots,D_K^t \right ]^T $ represents the vector of virtual queues for all IoVT devices in the $t$-th time slot.
It is evident that the size of the quadratic Lyapunov function $L\left ( \mathbf{D}^t \right )$ directly corresponds to the total length of the virtual queues.
Hence, as long as we can ensure that the Lyapunov function changes minimally between time slots, the stability of the virtual queue lengths can be guaranteed \cite{CuiA2012}.

To capture the variations in virtual queue lengths between different time slots, we define the conditional Lyapunov drift $\alpha \left ( \mathcal{S}^t \right ) $ as follows:
\begin{equation}\label{conditional Lyapunov drift}
  \begin{aligned}
    \alpha \left ( \mathcal{S}^t \right ) = \mathbb{E}\left \{ L\left ( \mathbf{D}^{t+1} \right ) - L\left ( \mathbf{D}^t \right ) \left .  \right |  \mathcal{S}^t \right \},
  \end{aligned}
\end{equation}
where $\mathcal{S}^t = \left \{ D_k^t,d_k^t \right \}_{k \in \mathcal{K} }$ represents the network state in the $t$-th time slot.
$\alpha \left ( \mathcal{S}^t \right )$ represents the expected change in the Lyapunov function between the $t$-th and $t+1$-th time slots, which can demonstrate the stability of the virtual queues.
Based on the drift-plus-penalty minimization method, we can obtain the optimal power control and IRS reflection matrix design scheme of the problem (\ref{orignal optimization problem}) in the $t$-th time slot by solving the following problem
\begin{subequations}\label{optimization problem 1}
\begin{align}
  \min_{\mathbf{v}^t, \mathbf{p}^t}\ \alpha \left ( \mathcal{S}^t \right ) + V\mathbb{E}\left \{ P^t \left .  \right | \mathcal{S}^t \right \}    \ \ \label{objection function 2}
\end{align}
\begin{alignat}{2}
\text{s.t.}\
             &(\ref{orignal constrited 1}),(\ref{orignal constrited 2}),  \label{1 constrited 1}
\end{alignat}
\end{subequations}
where $V$ represents the control parameter. By adjusting the control parameter $V$, we can achieve a trade-off between power consumption and queue stability. Specifically, a larger $V$ indicates that problem (\ref{optimization problem 1}) places more emphasis on power consumption, while a smaller $V$ places more emphasis on queue stability.

Due to the non-convexity introduced by the $\max \{ \cdot \}$ function in (\ref{queue update}) and (\ref{virtual queue update}), directly solving problem (\ref{optimization problem 1}) becomes highly computational complexity \cite{MoltafetPower2022}. To address this challenge, we turn to establish an upper bound for the objective function (\ref{objection function 2}).
The upper bound for the $\left ( D_k^{t+1} \right )^2$ can be expressed as
\begin{equation}\label{upper bound of virtual queue}
  \begin{aligned}
    \left ( D_k^{t+1} \right )^2 &= \left ( \max \left \{ D_k^{t} - d_k^{th} + d_k^{t+1},0 \right \} \right )^2 \\
                                 &\le \left ( D_k^{t} \right )^2  + \left ( d_k^{th} \right )^2  + \left ( d_k^{t+1} \right )^2 + 2D_k^{t}d_k^{t+1}, \forall k\in\mathcal{K}.
  \end{aligned}
\end{equation}
According to (\ref{upper bound of virtual queue}), we can obtain the upper bound to conditional Lyapunov drift $\mathcal{S}^t$ as
\begin{equation}\label{upper bound to Lyapunov function}
  \begin{aligned}
    &\alpha \left ( \mathcal{S}^t \right ) \le \frac{1}{2} \mathbb{E}\left \{ {\textstyle \sum_{k \in \mathcal{K}  }} \left ( \left ( d_k^{th} \right )^2  + \left ( d_k^{t+1} \right )^2 + 2D_k^{t}d_k^{t+1} \right ) | \mathcal{S}^t\right \}\\
                                          &= \frac{1}{2}{\textstyle \sum_{k \in \mathcal{K}  }}\left ( \left ( d_k^{th} \right )^2 - \left ( D_k^{t} \right )^2 + \mathbb{E}\left \{ \left ( d_k^{t+1} + D_k^{t} \right )^2  | \mathcal{S}^t\right \}  \right ).
  \end{aligned}
\end{equation}
Please note that when we replace $\mathcal{S}^t$ with its upper bound and minimize this upper bound, we can still ensure the stability of the virtual queue.
Then, we need to determine $\mathbb{E}\left \{ \left ( d_k^{t+1} + D_k^{t} \right )^2  | \mathcal{S}^t\right \}$ in (\ref{upper bound to Lyapunov function}). To this end, we impose the constraint $R_k^t  \tau \le Q_k^t +A_k^t$, then $Q_k^{t+1}$ equals to $Q_k^t +A_k^t - R_k^t \tau$. This constraint is justified because our objective is to minimize the long-term total power consumption under the delay constraint. When $R_k^t \tau \ge Q_k^t +A_k^t$, the queue delay remains the same as when $R_k^t \tau = Q_k^t +A_k^t$, but with higher power consumption. Based on the above analysis, under the constraint $R_k^t \tau \le Q_k^t +A_k^t$, the expression for $\mathbb{E}\left \{ \left ( d_k^{t+1} + D_k^{t} \right )^2  | \mathcal{S}^t\right \}$ can be represented as
\begin{equation}\label{expression of expection}
  \begin{aligned}
    &\mathbb{E}\left \{ \left ( d_k^{t+1} + D_k^{t} \right )^2  | \mathcal{S}^t\right \}  \\
    &\quad =\left ( \tilde{A}_k^{t} \right )^{-2} \mathbb{E}\left \{ \left ( Q_k^t +A_k^t - R_k^t \tau  + \tilde{A}_k^{t}D_k^{t} \right )^2  | \mathcal{S}^t \right \}\\
    &\quad= \left ( \tilde{A}_k^{t} \right )^{-2}\mathbb{E}\left \{  \left ( Q_k^t +A_k^t + \tilde{A}_k^{t}D_k^{t} \right )^2 \right .\\
    &\quad \quad \left . + \left ( R_k^t \right )^2  \tau^2 - 2 \left ( Q_k^t +A_k^t + \tilde{A}_k^{t}D_k^{t} \right ) R_k^t \tau   | \mathcal{S}^t \right \}\\
    &\quad \le U_k^t - \mathbb{E}\left \{  2\omega_k^t R_k^t   | \mathcal{S}^t \right \},\\
  \end{aligned}
\end{equation}
where $\omega_k^t = \left ( \tilde{A}_k^{t} \right )^{-2}\left ( Q_k^t +A_k^t + \tilde{A}_k^{t}D_k^{t} \right )\tau$. Besides, $U_k^t$ is obtained by
\begin{equation}\label{Uk in the equaility}
  \begin{aligned}
     &\left ( \tilde{A}_k^{t} \right )^{-2}\left ( \left ( Q_k^t +A_k^t + \tilde{A}_k^{t}D_k^{t} \right )^2 + \left ( R_k^t \right )^2 \tau^2 \right ) \le\\
     & \left ( \tilde{A}_k^{t} \right )^{-2}\left ( \left ( Q_k^t +A_k^t + \tilde{A}_k^{t}D_k^{t} \right )^2 + \left ( R_{k,max}^t \right )^2 \tau^2 \right ) \triangleq U_k^t,
  \end{aligned}
\end{equation}
where $R_{k,max}^t = B\log_2 \left (1+p_{k,max}\left | h_k^t \right |^2/\sigma^2\right )$ is the maximum data rate from the $k$-th IoVT device to the BS in the $t$-th time-slot.

According to (\ref{expression of expection}) and (\ref{upper bound to Lyapunov function}), the upper bound for objection function (\ref{objection function 2}) can be expressed as
\begin{equation}\label{upper bounder of OB}
  \begin{aligned}
  &\alpha \left ( \mathcal{S}^t \right ) + V\mathbb{E}\left \{ P^t \left .  \right | \mathcal{S}^t \right \}\\
  &\quad \le \mathbb{E}\left \{ VP^t  - {\textstyle \sum_{k \in \mathcal{K}  }}\left ( \omega_k^t R_k^t  \right ) \left .  \right | \mathcal{S}^t \right \} \\
  &\quad  + \frac{1}{2}{\textstyle \sum_{k \in \mathcal{K} }}\left ( \left ( d_k^{th} \right )^2 - \left ( D_k^{t} \right )^2  + U_k^t\right )
  \end{aligned}
\end{equation}
Note that (\ref{upper bounder of OB}) is obtained with the constraint $R_k^t \tau \le Q_k^t +A_k^t$.
Then we replace the objective function (\ref{objection function 2}) of the problem with (\ref{upper bounder of OB}). At each time slot, we utilize the stochastic minimization of the expected value method to solve the problem, which allows us to disregard the expectation in the objective function. Mathematically, the optimization problem in the $t$-th time slot can be formulated as
\begin{subequations}\label{optimization problem 3}
\begin{align}
  \min_{\mathbf{v}^t, \mathbf{p}^t}\  -  \sum_{k \in \mathcal{K}  } \omega_k^t R_k^t  + VP^t    \ \ \label{objection function 3}
\end{align}
\begin{alignat}{2}
\text{s.t.}\
             &R_k^t \tau \le Q_k^t +A_k^t,\ \forall k \in \mathcal{K},  \label{3 constrited 1}\\
             &(\ref{orignal constrited 1}),(\ref{orignal constrited 2}),   \label{3 constrited 2}
\end{alignat}
\end{subequations}
where the constant term in the objection function is omitted. Then, the primary steps of the proposed dynamic jointly optimization algorithm are summarized in Algorithm \ref{Algorithm1}.
At the beginning of each time slot, the BS observes the channel states and network states. Subsequently, leveraging the opportunistic expectation minimizing approach, the BS solves the problem (\ref{optimization problem 3}) to derive the optimal values of $\mathbf{v}^t$ and $\mathbf{p}^t$. After employing the optimal solution for uplink video data transmission, the network state is finally updated.

\begin{algorithm}[htbp]
	\caption{Dynamic Jointly Optimization Algorithm Based on Lyapunov Optimization}\label{Algorithm1}
	\KwIn{Set time slot $t = 1$, parameter V, and initialize $\left \{ Q_k^1 = 0, A_k^1 = 0\right \}_{k \in \mathcal{K} }$. }
	\While{$t \le T$}{
           Obtain the optimal IRS reflection matrix $\mathbf{v}^t$ and power control vector $\mathbf{p}^t$ by solving problem (\ref{optimization problem 3});

           With the $\mathbf{v}^t$ and $\mathbf{p}^t$, update queue state $\left \{ Q_k^{t+1} \right \}_{k \in \mathcal{K}} $ using (\ref{queue update});

           Set $t = t + 1$;
        }	
\end{algorithm}

However, it is worth noting that solving the non-convex problem (\ref{optimization problem 3}) can be challenging, primarily owing to the inherent discreteness of the phase-shift coefficients. Employing an exhaustive search method for its solution incurs substantial computational complexity and the coupling between power control and passive reflection matric design. Consequently, we introduce a tractable algorithm in the following section to solve the problem (\ref{optimization problem 3}) within affordable complexity.

\section{Problem Solution}
To solve the problem (\ref{optimization problem 3}) within affordable complexity, we propose a tractable algorithm that employs the alternating optimization to separately and iteratively solve $\mathbf{p}^t$ and $\mathbf{v}^t$. We decompose the original problem into two independent subproblems, namely uplink power control with fixed IRS PBF and the IRS reflection matrix design with fixed transmit power.
Through each iterative step, the ongoing process effectively reduces the objective function (\ref{objection function 3}), progressively resulting in the solution of problem (\ref{optimization problem 3}) towards an ultimate convergence to its optimal value.
In the rest of this section, we provide a detailed description of the proposed algorithm.

\subsection{Uplink Power Control}
When IRS reflecting matrix $\mathbf{v}^t$ is fixed, problem (\ref{optimization problem 3}) can be transformed into the power control subproblem as
\begin{subequations}\label{optimization problem 4}
\begin{align}
  \min_{\mathbf{p}^t}\  f_1\left ( \mathbf{p}^t \right ) =  \sum_{k \in \mathcal{K}  } \left ( - \omega_k^t R_k^t   + Vp_k^t  \right )    \ \ \label{objection function 4}
\end{align}
\begin{alignat}{2}
\text{s.t.}\
             &(\ref{orignal constrited 2}),(\ref{3 constrited 1}),   \label{4 constrited 1}
\end{alignat}
\end{subequations}
where the constant term in the objection function (\ref{objection function 4}) is omitted for simplicity. Then, the first-order derivative of $f_1\left ( \mathbf{p}^t \right )$ w.r.t $p_k^t$ is given by
\begin{equation}\label{first order derivative of f1}
  \begin{aligned}
    \frac{\mathrm{d} f_1\left ( \mathbf{p}^t \right )}{\mathrm{d} p_k^t} = \frac{-w_k^tB\left | h_k^t \right |^2 }{\ln2\left ( \sigma^2+p_k^t\left | h_k^t \right |^2  \right ) } +V.
  \end{aligned}
\end{equation}
Furthermore, we can obtain the second derivative of $f_1\left ( \mathbf{p}^t \right )$ w.r.t $p_k^t$ as following
\begin{equation}\label{second order derivative of f1}
  \begin{aligned}
    \frac{\mathrm{d}^2 f_1\left ( \mathbf{p}^t \right )}{\mathrm{d} \left ( p_k^t \right )^2 } = \frac{w_k^tB\left | h_k^t \right |^4 }{\ln2\left ( \sigma^2+p_k^t\left | h_k^t \right |^2  \right )^2 } \ge 0.
  \end{aligned}
\end{equation}
According to (\ref{second order derivative of f1}), function $f_1\left ( \mathbf{p}^t \right )$ is obviously convex. Therefore, the local minimizer of function $f_1\left ( \mathbf{p}^t \right )$ can be obtain by setting (\ref{first order derivative of f1}) equals to zero, which can be expressed as
\begin{equation}\label{local minimizer}
  \begin{aligned}
    p_{k,1}^t = \frac{\omega_k^tB}{V\ln2} -\frac{\sigma^2}{\left | h_k^t \right |^2 }, \ \forall k \in \mathcal{K}.
  \end{aligned}
\end{equation}
Besides, constraint (\ref{3 constrited 1}) limits the maximum data rate of each IoVT device. Since the data rate monotonically increases with transmit power, we can obtain the constraint transmit power of each IoVT device based on the constraint (\ref{3 constrited 1}), which is given by
\begin{equation}\label{transmit power limit}
  \begin{aligned}
    p_{k,2}^t = \frac{\sigma^2}{\left | h_k^t \right |^2 } \left ( 2^{\frac{Q_k^t +A_k^t}{B\tau}}-1 \right ), \ \forall k \in \mathcal{K}.
  \end{aligned}
\end{equation}
Based on the analysis of (\ref{local minimizer}) and (\ref{transmit power limit}), while considering the maximum transit power constraint (\ref{orignal constrited 2}), the optimal transmit power is obtained by
\begin{equation}\label{optimal transmit power}
  \begin{aligned}
    p_k^{t,opt} = \min\left \{ \max\left \{ 0,p_{k,1}^t \right \} ,p_{k,2}^t,p_{k,max} \right \}, \ \forall k \in \mathcal{K}.
  \end{aligned}
\end{equation}

\subsection{IRS Reflection Matrix Design}
When power control scheme $\mathbf{p}^t$ is fixed, problem (\ref{optimization problem 3}) can be transformed into the IRS reflection matrix design subproblem as
\begin{subequations}\label{optimization problem 5}
\begin{align}
  \max_{\mathbf{v}^t}\   f_{2}\left ( \mathbf{v}^t \right ) =   \sum_{k \in \mathcal{K}  }\tilde{\omega} _k^t\log_2\left ( 1 + \gamma_k^t \right )     \ \ \label{objection function 5}
\end{align}
\begin{alignat}{2}
\text{s.t.}\
             &(\ref{orignal constrited 1}),(\ref{3 constrited 1}),  \label{5 constrited 1}
\end{alignat}
\end{subequations}
where $\tilde{\omega} _k^t = \omega _k^t B$ and $\gamma_k^t = \frac{p_k^t\left | h_k^t \right |^2 }{\sigma^2}$.
Directly solving the sum-of-logarithms-of-ratio problem (\ref{optimization problem 5}) involves high complexity. To ensure low complexity, the closed-form FP method proposed in \cite{ShenFractional2018} is an effective solution.
The details of solving the problem using the FP method are described in the following.

We begin by employing the Lagrange dual transformation to handle the logarithmic form of the objective function (\ref{objection function 5}). By introducing an auxiliary variables $\pmb{\eta}^t = \left [\eta_1^t,\cdots,\eta_K^t  \right ]^T $, the objective function (\ref{objection function 5}) can be transformed based on the following equation
\begin{equation}\label{LDT equation}
  \begin{aligned}
    \log_2\left ( 1+\gamma_k^t \right ) =\max_{\eta_k^t\ge 0}\  \log_2\left ( 1+\eta_k^t \right ) - \eta_k^t + \frac{\left ( 1+\eta_k^t \right )\gamma_k^t }{1+\gamma_k^t}.
  \end{aligned}
\end{equation}
Based on (\ref{LDT equation}), the problem (\ref{optimization problem 5}) can be equivalently reexpressed as
\begin{subequations}\label{optimization problem add}
\begin{align}
  \max_{\mathbf{v}^t,\pmb{\eta}^t}  &\sum_{k \in \mathcal{K}}\tilde{\omega} _k^t\log_2\left ( 1+\eta_k^t \right ) - \sum_{k \in \mathcal{K}}\tilde{\omega} _k^t\eta_k^t + \sum_{k \in \mathcal{K}}\frac{\tilde{\omega} _k^t\left ( 1+\eta_k^t \right )\gamma_k^t }{1+\gamma_k^t}   \ \ \label{objection function add}
\end{align}
\begin{alignat}{2}
\text{s.t.}\
             & \eta_k^t \ge 0,\ \forall k \in \mathcal{K},  \label{add constrited 1}\\
             & (\ref{orignal constrited 1}), (\ref{3 constrited 1}). \label{add constrited 2}
\end{alignat}
\end{subequations}
By solving for $\mathbf{v}^t$ and $\pmb{\eta}^t$ alternately, we can obtain the solution for the problem (\ref{optimization problem add}). Specifically, when $\mathbf{v}^t$ is fixed, the optimal $\eta_k^t$ is obtained by
\begin{equation}\label{optimal eta}
  \begin{aligned}
    \eta_k^{t,opt} = \gamma_k^t,\ \forall k \in \mathcal{K}.
  \end{aligned}
\end{equation}
With fixed $\pmb{\eta}^t$, we focus on the following sum-of-ratios problem
\begin{subequations}\label{optimization problem 51}
\begin{align}
  \max_{\mathbf{v}^t}\   f_{3a}\left ( \mathbf{v}^t \right ) =  \sum_{k \in \mathcal{K}} \frac{\tilde{\omega} _k^t\left ( 1+\eta_k^t \right )\gamma_k^t }{1+\gamma_k^t}     \ \ \label{objection function 51}
\end{align}
\begin{alignat}{2}
\text{s.t.}\
             &(\ref{orignal constrited 1}),(\ref{3 constrited 1}).  \label{51 constrited 1}
\end{alignat}
\end{subequations}
For the purpose of a clear description, based on (\ref{represent of overall channel}), we reexpress the objection function $f_{3a}\left ( \mathbf{v}^t \right )$ as
\begin{equation}\label{reexpression of f4a}
  \begin{aligned}
    f_{3b}\left ( \mathbf{v}^t \right ) = \sum_{k \in \mathcal{K}}\frac{\tilde{\eta} _k^tp_k^t\left | h_{d,k}^t + \left ( \mathbf{v}^t \right )^H \mathbf{h}_{c,k}^t \right |^2 }{ p_k^t\left | h_{d,k}^t + \left ( \mathbf{v}^t \right )^H \mathbf{h}_{c,k}^t \right |^2 + \sigma^2},
  \end{aligned}
\end{equation}
where $\tilde{\eta} _k^t = \tilde{\omega} _k^t\left ( 1+\eta_k^t \right )$.
To solve the non-convex problem (\ref{optimization problem 51}), we employ the quadratic transformation technique for handling the objection function (\ref{reexpression of f4a}).
By introducing the auxiliary variables $\pmb{\zeta}^t = \left [ \zeta_1^t,\cdots,\zeta_K^t \right ] $, and the problem (\ref{optimization problem 51}) can be equivalently reexpressed as
\begin{subequations}\label{optimization problem 52}
\begin{align}
  \max_{\mathbf{v}^t,\pmb{\zeta}^t}\   &\sum_{k \in \mathcal{K}} 2 \sqrt{\tilde{\eta} _k^tp_k^t} \text{Re}\left \{ \left ( \zeta_k^t \right )^* \left ( h_{d,k}^t + \left ( \mathbf{v}^t \right )^H \mathbf{h}_{c,k}^t \right )  \right \} \nonumber \\
                                                           &- \sum_{k \in \mathcal{K}} \left | \zeta_k^t \right |^2\left ( p_k^t\left | h_{d,k}^t + \left ( \mathbf{v}^t \right )^H \mathbf{h}_{c,k}^t \right |^2 + \sigma^2 \right )        \ \ \label{objection function 52}
\end{align}
\begin{alignat}{2}
\text{s.t.}\
             &(\ref{orignal constrited 1}),(\ref{3 constrited 1}).  \label{52 constrited 1}
\end{alignat}
\end{subequations}
The problem (\ref{optimization problem 52}) can also be solved by updating $\mathbf{v}^t$ and $\pmb{\zeta}^t$ alternatively.
Specifically, while keeping $\mathbf{v}^t$ constant, we first equate the first-order derivative of the objective function (\ref{objection function 52}) to zero with respect to $\pmb{\zeta}^t$ and the optimal $\zeta_k^t$, $\forall k \in \mathcal{K}$ can be calculated as
\begin{equation}\label{optimal yk}
  \begin{aligned}
    \zeta_k^{t,opt} = \frac{\sqrt{\tilde{\eta} _k^tp_k^t}\left ( h_{d,k}^t + \left ( \mathbf{v}^t \right )^H \mathbf{h}_{c,k}^t \right )}{p_k^t\left | h_{d,k}^t + \left ( \mathbf{v}^t \right )^H \mathbf{h}_{c,k}^t \right |^2 + \sigma^2}.
  \end{aligned}
\end{equation}
Then, we update $\mathbf{v}^t$ with fixed $\pmb{\zeta}^t$. For the purpose of a clear description, the $\left | h_{d,k}^t + \left ( \mathbf{v}^t \right )^H \mathbf{h}_{c,k}^t \right |^2$ can be reexpressed as
\begin{equation}\label{open expression}
  \begin{aligned}
    &\left | h_{d,k}^t + \left ( \mathbf{v}^t \right )^H \mathbf{h}_{c,k}^t \right |^2 =  \left ( \mathbf{v}^t \right )^H \mathbf{h}_{c,k}^t\left ( \mathbf{h}_{c,k}^t\right )^H \mathbf{v}^t\\
    & \qquad \qquad  + 2\text{Re}\left \{ \left ( h_{d,k}^t\right )^*\left ( \mathbf{v}^t \right )^H \mathbf{h}_{c,k}^t  \right \}  + \left | h_{d,k}^t \right | ^2.
  \end{aligned}
\end{equation}
According to (\ref{open expression}), we can reexpress the objection function (\ref{objection function 52}) and the problem (\ref{optimization problem 52}) with fixed $\pmb{\zeta}^t$ can be transformed into
\begin{subequations}\label{optimization problem 53}
\begin{align}
  \max_{\mathbf{v}^t}\   &f_{4a}\left ( \mathbf{v}^t \right ) =  -\left ( \mathbf{v}^t \right )^H \mathbf{W}^t \mathbf{v}^t + 2\text{Re}\left \{ \left ( \mathbf{v}^t \right )^H\mathbf{q}^t  \right \} + C         \ \ \label{objection function 53}
\end{align}
\begin{alignat}{2}
\text{s.t.}\
             &(\ref{orignal constrited 1}),(\ref{3 constrited 1}),  \label{53 constrited 1}
\end{alignat}
\end{subequations}
where
\begin{subequations}\label{explain of objection function}
  \begin{align}
    \mathbf{W}^t = \sum_{k \in \mathcal{K}}\left | \zeta_k^t \right |^2 p_k^t \mathbf{h}_{c,k}^t\left ( \mathbf{h}_{c,k}^t\right )^H,
  \end{align}
  \begin{align}
    \mathbf{q}^t = \sum_{k \in \mathcal{K}} \left ( \sqrt{\tilde{\eta} _k^tp_k^t}\left ( \zeta_k^t \right )^*\mathbf{h}_{c,k}^t - \left | \zeta_k^t \right |^2 p_k^t\left ( h_{d,k}^t\right )^* \mathbf{h}_{c,k}^t  \right ),
  \end{align}
  \begin{align}
    C = &\sum_{k \in \mathcal{K}} \left ( 2 \sqrt{\tilde{\eta} _k^tp_k^t} \text{Re}\left \{ \left ( \zeta_k^t \right )^*  h_{d,k}^t \right \} \right . \nonumber \\
        & \left .- \left | \zeta_k^t \right |^2\left ( p_k^t\left | h_{d,k}^t \right | ^2+ \sigma^2 \right )  \right ).
  \end{align}
\end{subequations}
Directly solving quadratic programming problem (\ref{explain of objection function}) with discrete variable constraints incurs a significant computational complexity \cite{ZhangA2021}. To address this challenge, we adopt an iterative optimization approach in which we sequentially optimize each element of $\mathbf{v}^t$ while keeping the remaining $MN - 1$ elements constant.
When solving the $n$-th element in $\mathbf{v}^t$, we only focus on the part of the objection function (\ref{objection function 53}) related to $v_n^t$. To make the portion of the objective function (\ref{objection function 53}) related to $v_n^t$ more explicit, we reexpress $\left ( \mathbf{v}^t \right )^H\mathbf{q}^t$ and $\left ( \mathbf{v}^t \right )^H \mathbf{W}^t \mathbf{v}^t$ as
\begin{equation}\label{open the expression with on variable1}
  \begin{aligned}
    \left ( \mathbf{v}^t \right )^H\mathbf{q}^t = \sum_{i=1}^{MN}\left ( v_i^t \right )^*q_i^t = \left ( v_n^t \right )^*q_n^t + \sum_{i=1,i\ne n}^{MN} \left ( v_i^t \right )^*q_i^t,
  \end{aligned}
\end{equation}
\begin{equation}\label{open the expression with on variable2}
  \begin{aligned}
    &\left ( \mathbf{v}^t \right )^H \mathbf{W}^t \mathbf{v}^t = \sum_{i=1}^{MN}\sum_{j=1}^{MN}\left ( v_i^t \right )^* w_{i,j}^t v_j^t\\
                                                              &= \left ( v_n^t \right )^* w_{n,n}^t v_n^t + 2 \text{Re}\left \{ \sum_{j=1,j\ne n}^{MN}\left ( v_n^t \right )^* w_{n,j}^t v_j^t\right \}\\
                                                              &\quad +  \sum_{i=1,i\ne n}^{MN}\sum_{j=1,j\ne n}^{MN} \left ( v_i^t \right )^* w_{i,j}^t v_j^t.
  \end{aligned}
\end{equation}
Note that $\mathbf{W}^t$ is a hermitian matrix and then $w_{i,j}^t = \left ( w_{j,i}^t \right )^* $.
According to (\ref{open the expression with on variable1}) and (\ref{open the expression with on variable2}), we can reexpress the objection function (\ref{objection function 53}) to a function of $v_n^t$ as
\begin{equation}\label{reexpression of f5a}
  \begin{aligned}
    &f_{4b}\left ( v_n^t \right ) =  z\left ( v_1^t,\cdots,v_{n-1}^t,v_{n+1}^t,\cdots,v_{MN} \right ) \\
                                 &\quad - \left | v_n^t \right |^2w_{n,n}^t + 2 \text{Re}\left \{ \left ( v_n^t \right )^*\left ( q_n^t -  {\textstyle \sum_{j=1,j\ne n}^{MN}} w_{n,j}^tv_j^t  \right )   \right \} .
  \end{aligned}
\end{equation}
Since $z\left ( v_1^t,\cdots,v_{n-1}^t,v_{n+1}^t,\cdots,v_{MN} \right )$ remains constant when changing $v_n^t$, it can be omitted when updating the $v_n^t$. Besides, $\left | v_n^t \right |^2 = 1$ and the second term of (\ref{reexpression of f5a}) can also be disregarded during the optimization of the $n$-th phase. Thus, we only focus on the third term of (\ref{reexpression of f5a}) when updating the $v_n^t$. Let us denote the argument of $v_n^t$, $q_n^t -  {\textstyle \sum_{j=1,j\ne n}^{MN}} w_{n,j}^tv_j^t$ by $\angle v_n^t$, $\angle d_n^t$, respectively. Note that $\angle v_n^t$ corresponds to the phase of the $n$-th element of the IRS equivalent phase matrix, i.e., $\angle v_n^t = \phi_{m,n'}^t$ and $n = (m-1)N + n'$, $m = 1,\cdots,M$, $n'=1,\cdots,N$. The optimal phase $\angle v_n^t$ can be calculated by
\begin{subequations}\label{optimization problem 54}
\begin{align}
  \angle v_{n}^{t,opt} = \arg \max_{\angle v_{n}^{t} \in \mathcal{X}} \cos\left ( \angle d_n^t -\angle v_n^t \right )       \ \ \label{objection function 54}
\end{align}
\begin{alignat}{2}
\text{s.t.}\
             &(\ref{3 constrited 1}).  \label{54 constrited 1}
\end{alignat}
\end{subequations}
Then, we can obtain all the elements of $\mathbf{v}^t$ through iterative optimization.

\subsection{Global Total Power Minimization}
\begin{algorithm}[htbp]
	\caption{Two-Step Alternating Optimization Algorithm for Problem (\ref{optimization problem 3})}\label{Algorithm2}
	\KwIn{The tolerance $\xi$, iteration number $i$ and $i_v$ that are both set to 1 and the upper bound $I_{out}$, $I_v$, a feasible solution $\mathbf{p}^t(0)$ and $\mathbf{v}^t(0)$ of problem (\ref{optimization problem 3}).}
	\While{$\left | \text{Obj}^{t}\left ( i \right ) - \text{Obj}^t\left ( i-1 \right )  \right |^2 \ge \xi $ and $i \le I_{out}$}{
           Update $\mathbf{p}^t(i)$ with fixed $\mathbf{v}^t(i-1)$ by using (\ref{optimal transmit power});

           Set $i_v = 1$;

           \Repeat{$\left | \mathbf{v}^{t}\left ( i_v \right ) - \mathbf{v}^t\left ( i_v-1 \right )  \right |^2 < \xi $ or $i_v > I_v$}{
           Update $\pmb{\zeta}^t\left ( i_v \right )$ according to (\ref{optimal yk}) (fixed $\mathbf{p}^t(i)$);

           Update $\pmb{\eta}^t\left ( i_v \right )$ according to (\ref{optimal eta}) (fixed $\mathbf{p}^t(i)$);

           Update $\mathbf{v}^t\left ( i_v \right )$ according to (\ref{optimization problem 54}) (fixed $\mathbf{p}^t(i)$);

           $i_v = i_v + 1$;
           }

           obtain $\mathbf{v}^{t}\left ( i \right )=\mathbf{v}^{t}\left ( i_v \right )$ with fixed $\mathbf{p}^t(i)$;

           Set $i = i + 1$;
        }	

	\KwOut{The optimal $\mathbf{p}^t(i)$ and $\mathbf{v}^t(i)$.}
\end{algorithm}

Based on the solutions for the power control subproblem (\ref{optimization problem 4}) and the IRS reflection matrix design subproblem (\ref{optimization problem 5}), we propose an algorithm to address the problem (\ref{optimization problem 3}), as illustrated in Algorithm \ref{Algorithm2}.
Specifically, we initiate the power control scheme and IRS reflection matrix at the outset. Subsequently, in the $i$-th iteration, we individually calculate the optimal power control scheme and employ the FP method to obtain the IRS reflection matrix.
Define the value of objection function (\ref{objection function 3}) after the $i$-th iteration as $\text{Obj}^{t}(i)$.
The optimization of both $\mathbf{p}^t$ and $\mathbf{v}^t$ results in an increase in the objective function (\ref{objection function 3}) in the $i$-th iteration, meaning that $\text{Obj}^{t}(i) \ge \text{Obj}^{t}(i-1)$.
Furthermore, due to the constraints on achievable rates and transmission power in the problem (\ref{optimization problem 3}), the feasible solution space for the problem (\ref{optimization problem 3}) is limited.
Therefore, after a finite number of iterations, Algorithm \ref{Algorithm2} will converge.

\textbf{Computational Complexity Analysis:} Algorithm \ref{Algorithm2} is a two-step alternating optimization algorithm; therefore, its complexity is primarily determined by the complexity of its subproblems. For the power control subproblem (\ref{optimization problem 4}), its computational complexity is determined by the number of IoVT devices, of which the complexity is $\mathcal{O}\left ( K \right ) $. As for the IRS reflection design subproblem (\ref{optimization problem 5}), its computational complexity is primarily determined by the number of iterations $I_v$ and the computational complexity for each iteration. At each iteration, we alternately solve the all $MN$ elements in $\mathbf{V}^t$, which leads to a complexity of $\mathcal{O}\left ( L^2N^2 \right )$.
Assuming the maximum number of iterations for Algorithm \ref{Algorithm2} is $I_{out}$, its upper-bound computational complexity can be given by $ \mathcal{O}\left (I_{out}\left (K+ I_v\left ( 2K + L^2N^2 \right ) \right )  \right ) $.

\section{Performance Evaluation}
In this section, extensive numerical results are provided to validate the effectiveness of the proposed long-term total power minimization algorithm. In summary, we conduct simulations of a multi-IRS assisted IoVT system to assess the performance improvement attained by the proposed algorithm.

\subsection{Channel Model}
All channels are modeled as Rician fading channel in the proposed multi-IRS assisted IoVT system. Thus, the BS-IRS channel from BS to the $m$-th IRS in the $t$-th time-slot can be expressed as
\begin{equation}\label{channel between IRS }
  \mathbf{h}_{m}^t = PL^t_{m}\left ( \sqrt{\frac{\varepsilon_{BR}}{\varepsilon_{BR}+1}}\mathbf{h}_{m}^{LOS,t}+\sqrt{\frac{1}{\varepsilon_{BR}+1}}\mathbf{h}_{m}^{NLOS,t}  \right ),
\end{equation}
where $PL^t_{m}$ denotes the path loss from the BS to the $m$-th IRS in the $t$-th time slot, $\varepsilon_{BR}$ is the Rician factor of $\mathbf{h}_{m}^t$, $\mathbf{h}_{m}^{LOS,t}$ and $\mathbf{h}_{m}^{NLOS,t}$ represent the LOS and NLOS components, respectively. Each element of $\mathbf{h}_{m}^{NLOS,t}$ follows a complex Gaussian distribution with zero mean and unit variance.
In particular, $PL^t_{m}$ is given by
\begin{equation}\label{path loss}
  \begin{aligned}
    PL^t_{m} = L_0 \left ( \frac{d_m}{D_0}  \right )^{-\iota_{BR}},
  \end{aligned}
\end{equation}
where $L_0$ denotes the path loss at the reference distance $D_0 = 1$ m, $d_m$ denotes the individual link distance between the BS and the $m$-th IRS, and $\iota_{BR}$ is the path loss exponent of BS-IRS channel. Besides, $\mathbf{h}_{m}^{LOS,t}$ is given by
\begin{equation}\label{expresson of LoS}
  \begin{aligned}
    \mathbf{h}_{m}^{LOS,t} = \mathbf{a}_x\left ( \varphi_{AoD}^{m,t} \right ) \otimes \mathbf{a}_y\left ( \vartheta_{AoD}^{m,t},\varphi_{AoD}^{m,t} \right ),
  \end{aligned}
\end{equation}
where $\vartheta_{AoD}^{m,t}$ and $ \varphi_{AoD}^{m,t}$ denotes the azimuth/elevation angle of departure for the link from the $m$-th IRS to the BS in the $t$-th time slot. $\mathbf{a}_x\left ( \varphi_{AoD}^{m,t} \right )$ and $\mathbf{a}_y\left ( \vartheta_{AoD}^{m,t},\varphi_{AoD}^{m,t} \right )$ are given by
\begin{subequations}\label{expression of AoD}
  \begin{align}
    \mathbf{a}_x\left ( \varphi_{AoD}^{m,t} \right ) = \left [1,e^{j\psi^{m,t}},\cdots,e^{j\left ( N_x -1 \right )\psi^{m,t} }\right ]^T,
  \end{align}
  \begin{align}
    \mathbf{a}_y\left ( \vartheta_{AoD}^{m,t},\varphi_{AoD}^{m,t} \right ) = \left [1,e^{j\chi^{m,t}} ,\cdots,e^{j\left ( N_y -1 \right ) \chi^{m,t}} \right ]^T,
  \end{align}
\end{subequations}
where $\psi^{m,t} = 2\pi\frac{d}{\lambda}\cos \varphi_{AoD}^{m,t}$, $\chi^{m,t} = 2\pi\frac{d}{\lambda}\sin \varphi_{AoD}^{m,t} \cos\vartheta_{AoD}^{m,t}$, $N_x$ and $N_y$ denote the number of IRS elements along the horizontal and vertical, respectively, $\lambda$ is the wavelength, $d$ is the antenna spacing. Note that each antenna spacing is half wavelength in the system. Since the BS-IRS channel distance is relatively moderate and randomly scattered, we set $\iota_{BR} = 2.2$ and $\varepsilon_{BR} = 1$. The IRS-device and BS-device channels are also generated by following the similar procedure. Based on this,
the IRS-device and BS-device channels are also dominated by LOS and NLOS links with the path loss exponent of 2.2 and 3.5, the Rician factor of 1 and 0.5, respectively. More details are omitted for the sake of brevity.

\subsection{Simulation Setup and Comparison Algorithms}
We use a 3D coordinate system to describe the system deployment. In detail, the coordinate of the BS is ($X_{BS} = -200$ m, $Y_{BS} = 0$, $Z_{BS} = 0$). All IRSs are located on the y-z plane with $X_{IRS} = 0$ and all IoVT devices are located on the x-z plane with $Y_{IRS} = 0$. The IRSs are located with equal spacing in a half circle with a diameter of 10 m and are facing the center of the coordinate. In addition, IoVT devices are randomly deployed within a circular area in the x-z plane with a center of ($0,200$ m) and a radius of 100 m.
There are $M = 2$ IRSs in the system, where each IRS comprises $N = N_xN_y$ elements. We fix $N_y = 4$ and increase $N_x$ linearly with $N$, the power consumption of each element is $P_{I,n} = 2$ dBm. The bandwidth of each sub-carrier is $15$ KHz and the noise power density is $-170$ dBm/Hz. Other required parameters are set as follows: $K = 10$, $\xi = 10^{-4}$, $\tau = 10$ ms, $b = 3$, $V = 50$, $d_k^{th} = 50$ ms, $p_{k,max} = 20$ dBm, $A_{k,min}^t = 1$ B, $A_{k,max}^t = 150$ B, $\forall k\in \mathcal{K}, t \in \mathcal{T}$.

To validate the effectiveness of our proposed algorithm, we compare their performance against the following benchmark algorithms:
\begin{itemize}
	
	\item  \textbf{Random IRS phase shift:} The phases in $\mathbf{v}$ are randomly selected, while the optimal power control scheme (\ref{optimal transmit power}) is performed.
	
	\item \textbf{Without IRS:} All IoVT devices transmit video data directly to BS without reflection from any IRS, and apply optimal power control scheme (\ref{optimal transmit power}).

    \item \textbf{Exhaustive search:} Traverse the solution space of problem (\ref{optimization problem 3}) at each time slot to find the optimal solution.
\end{itemize}

\subsection{Numerical Results and Performance Analysis}
\begin{figure}[ht]
	\centering
    \setlength{\abovecaptionskip}{-0.1cm}   
	\includegraphics[scale=0.37]{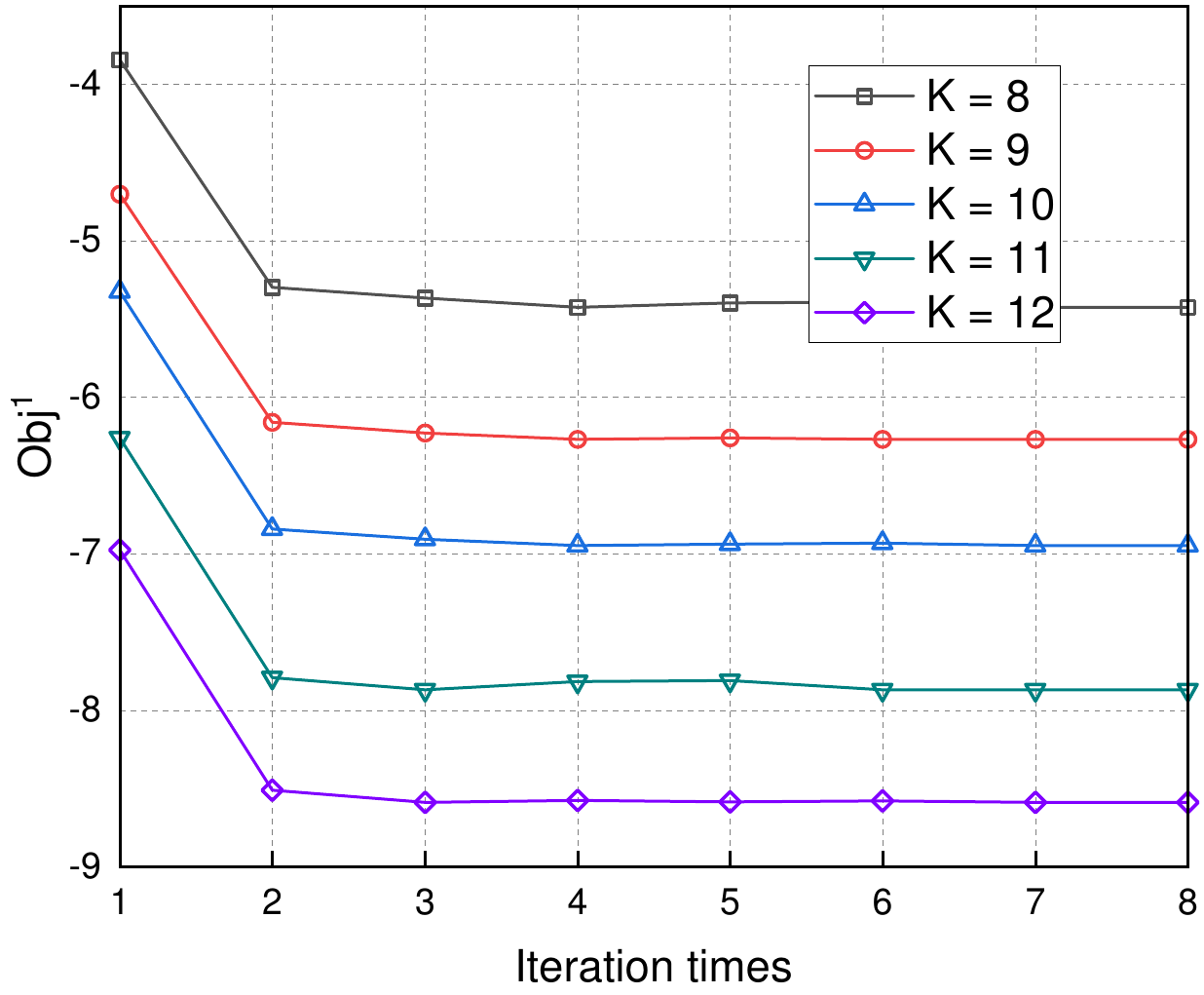}
    \caption{Convergence of the Proposed algorithm vs. number of the IoVT devices.}
	\label{convergencegraph}
\end{figure}

\begin{figure}[ht]
	\centering
    \setlength{\abovecaptionskip}{-0.1cm}   
	\includegraphics[scale=0.37]{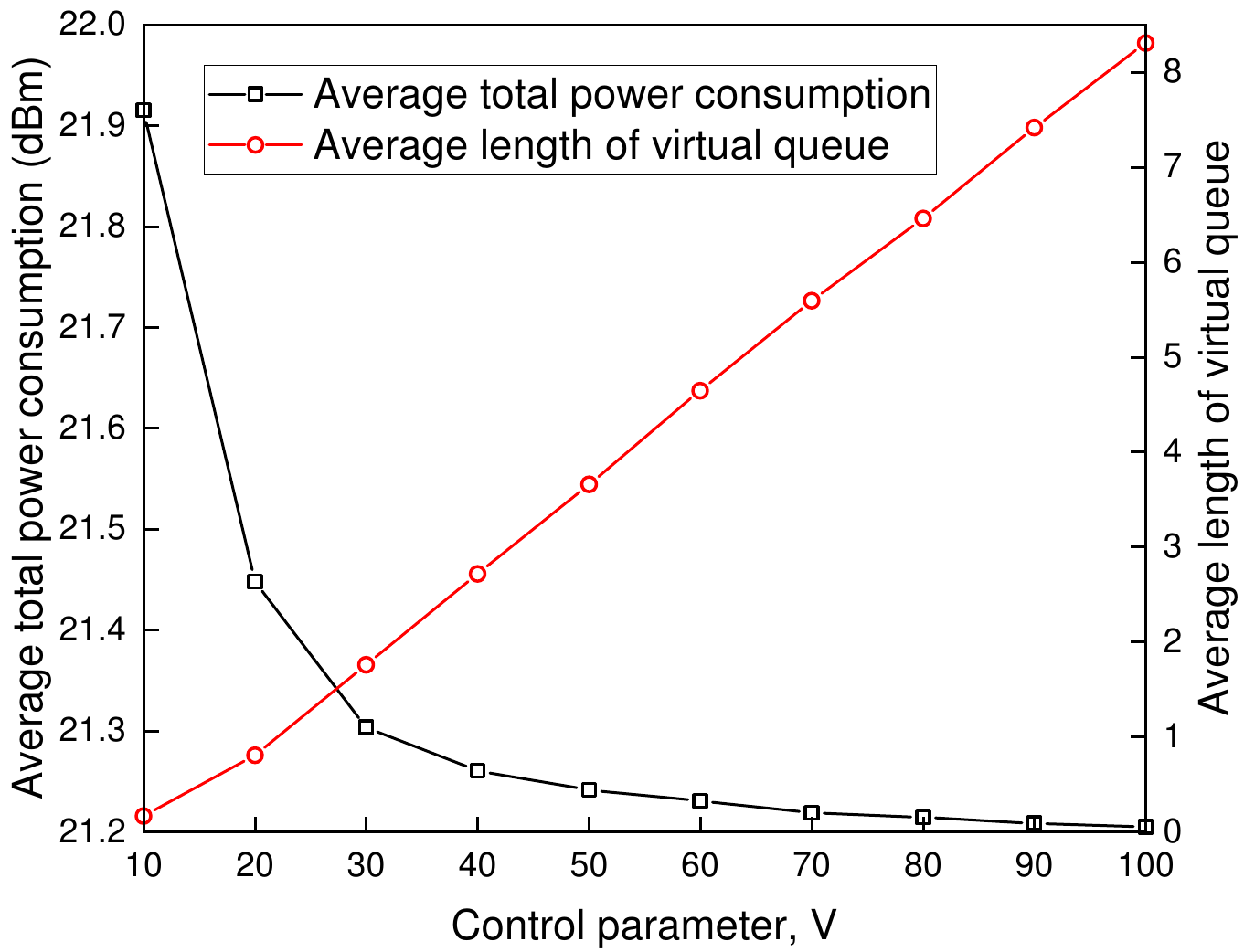}
    \caption{Trade-off of between the average total power consumption and average length of the virtual queues with different control parameter.}
	\label{PowerBlockVgraph}
\end{figure}

\begin{figure}[ht]
	\centering
    \setlength{\abovecaptionskip}{-0.1cm}   
	\includegraphics[scale=0.37]{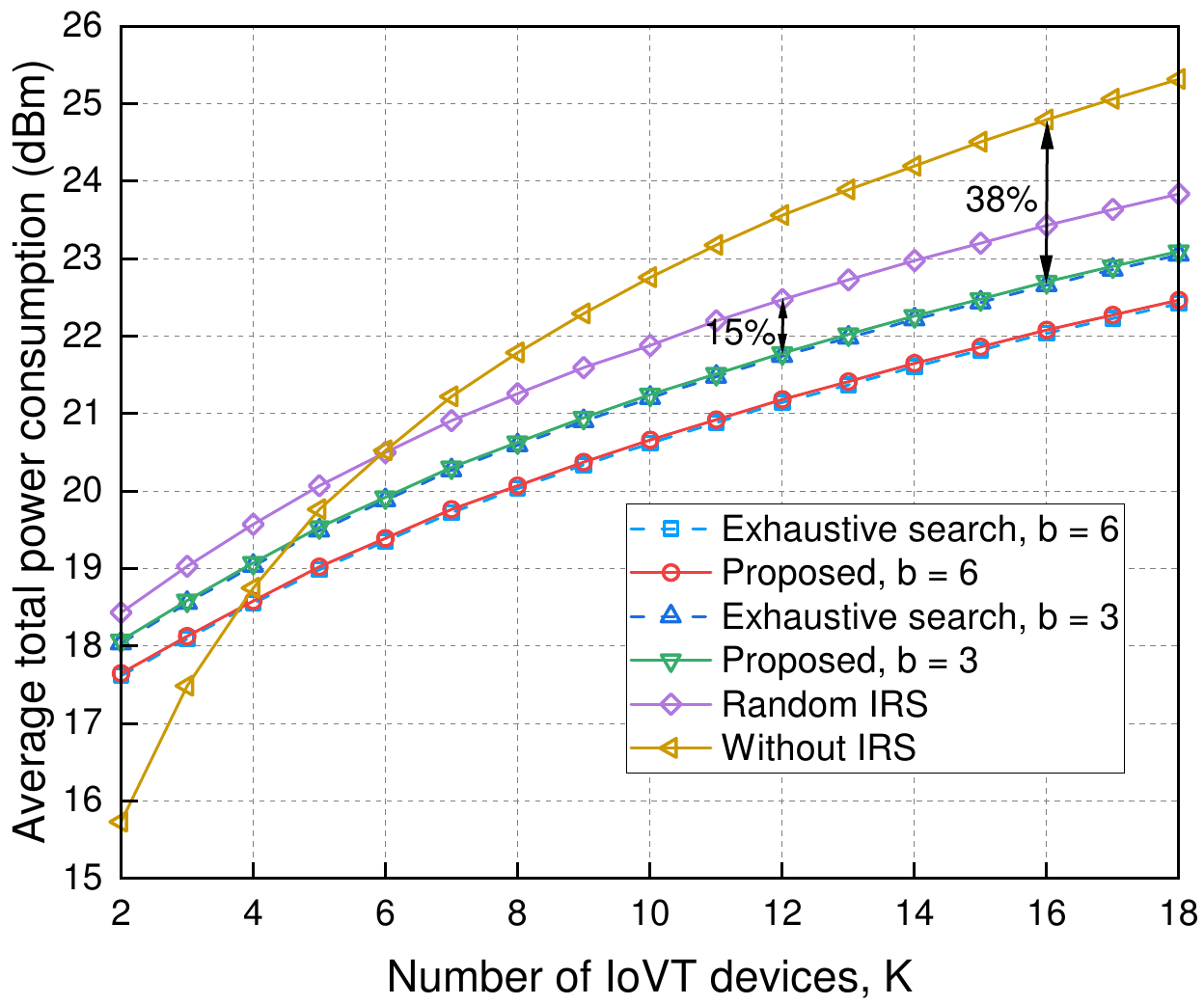}
    \caption{Average total power consumption vs. number of the IoVT devices.}
	\label{PowerKgraph}
\end{figure}

\begin{figure}[ht]
	\centering
    \setlength{\abovecaptionskip}{-0.1cm}   
	\includegraphics[scale=0.37]{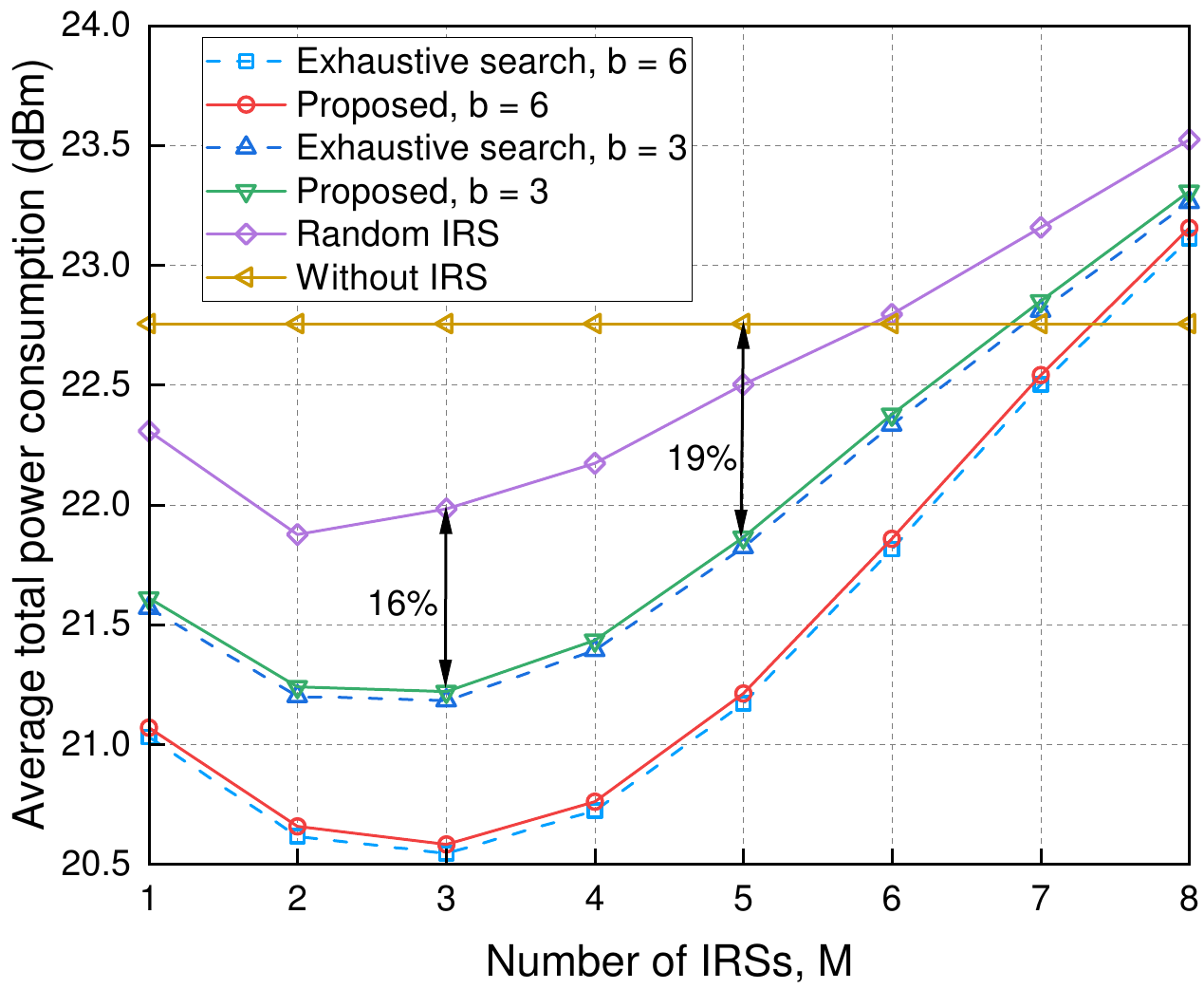}
    \caption{Average total power consumption vs. number of the IRSs.}
	\label{PowerMgraph}
\end{figure}

\begin{figure}[ht]
	\centering
    \setlength{\abovecaptionskip}{-0.1cm}   
	\includegraphics[scale=0.37]{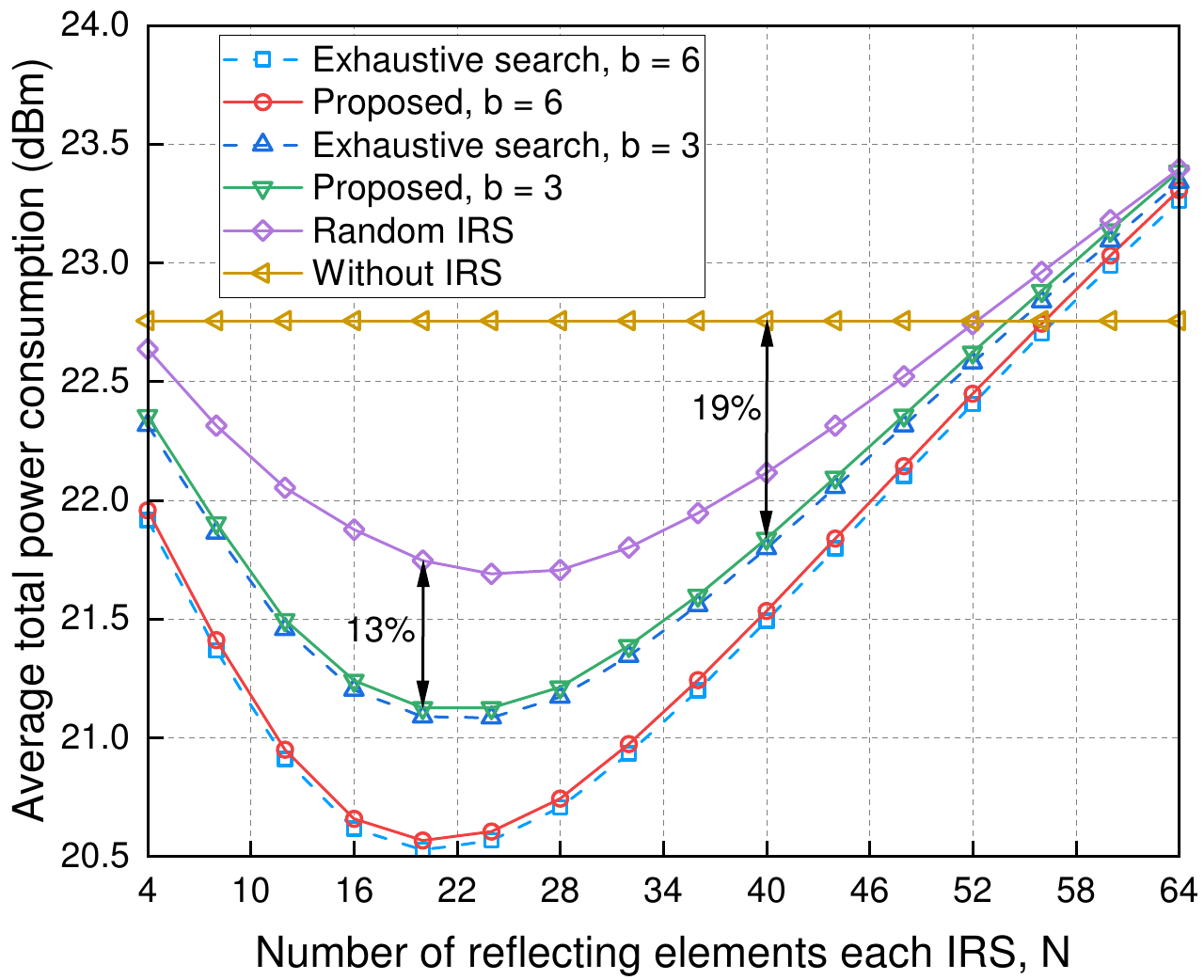}
    \caption{Average total power consumption vs. number of the reflecting elements each IRS.}
	\label{PowerNgraph}
\end{figure}

\begin{figure}[ht]
	\centering
    \setlength{\abovecaptionskip}{-0cm}   
	\includegraphics[scale=0.37]{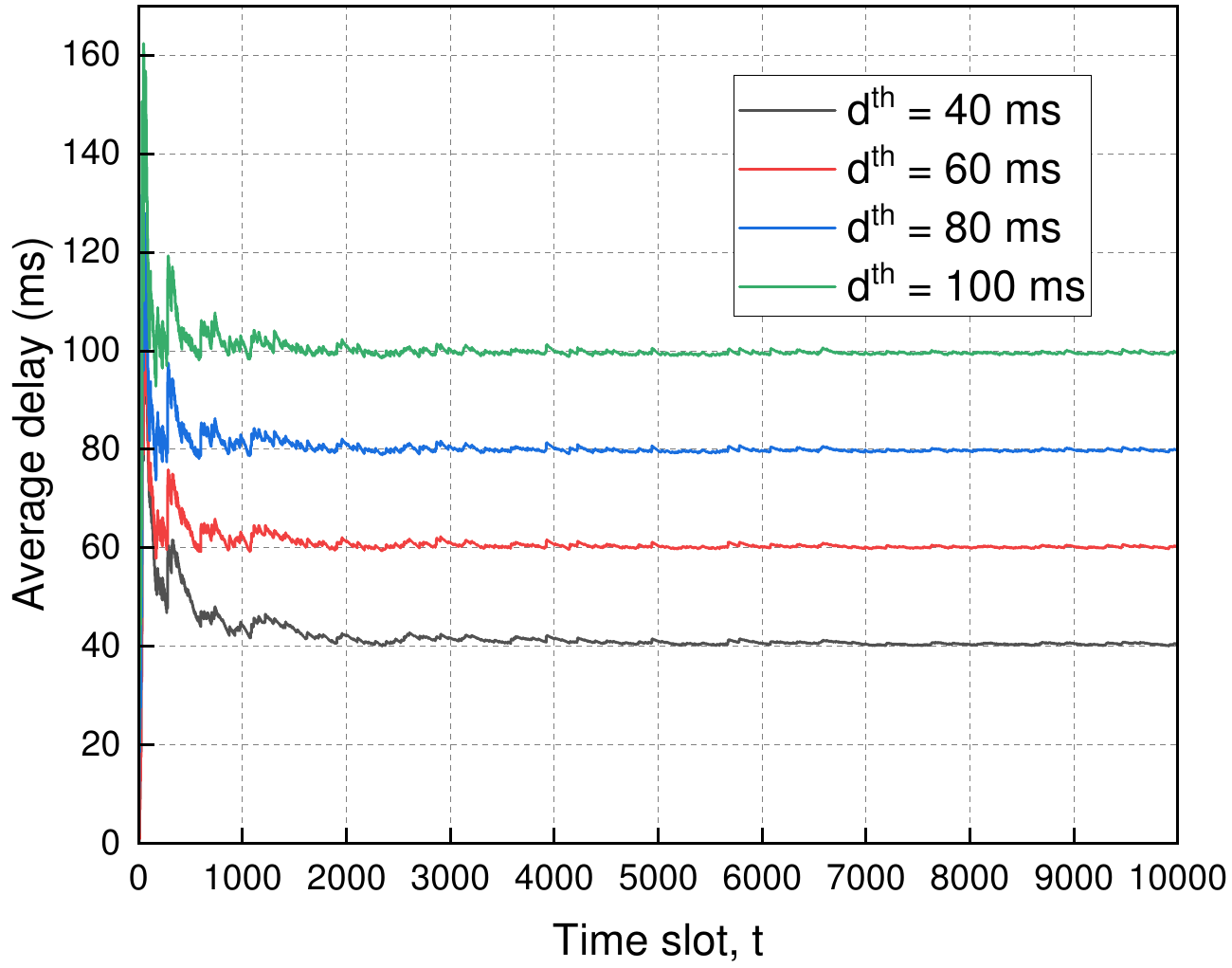}
    \caption{Evolution of the average delay with different delay threshold.}
	\label{DavTgraph}
\end{figure}

\begin{figure}[ht]
	\centering
    \setlength{\abovecaptionskip}{-0.1cm}   
	\includegraphics[scale=0.37]{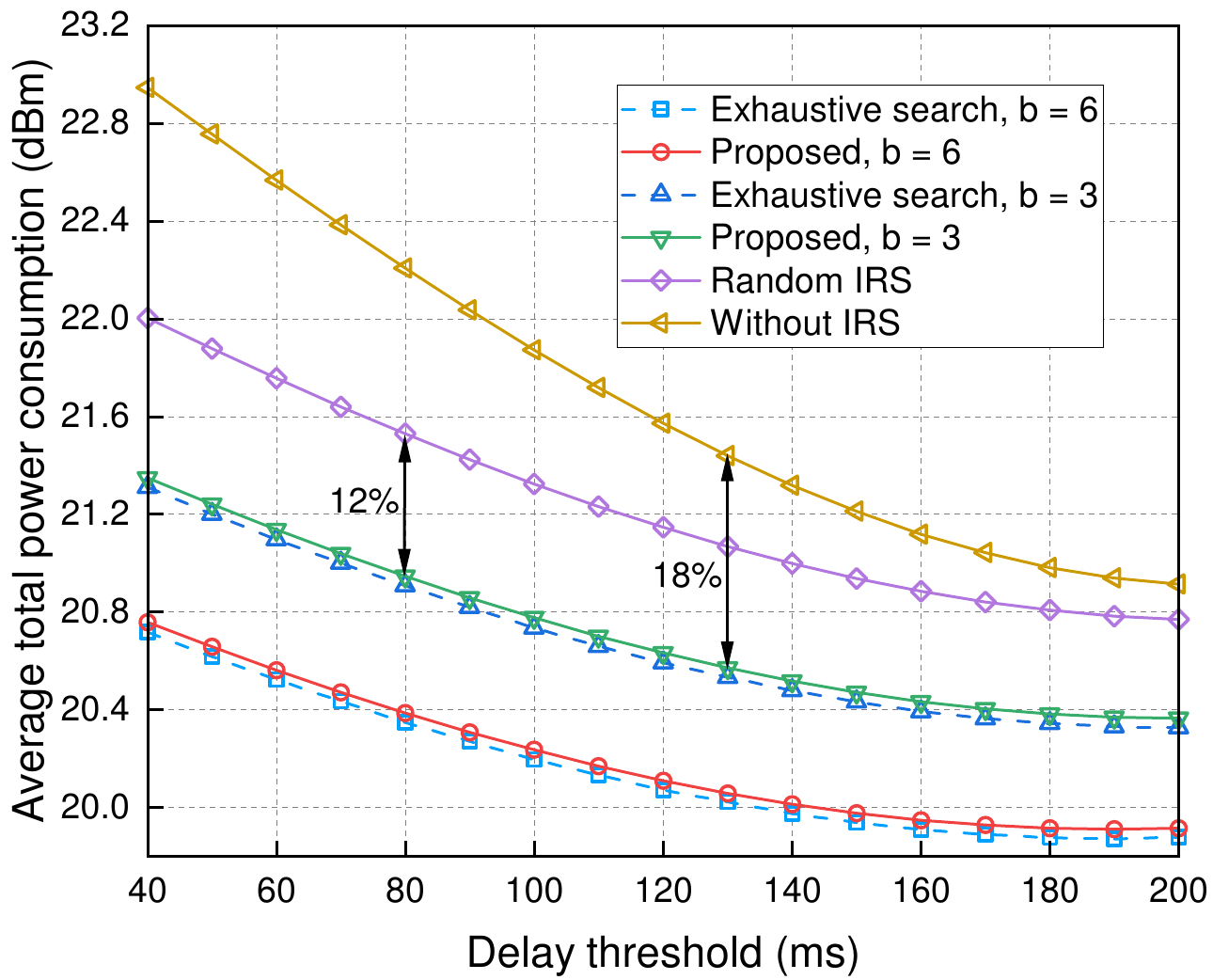}
    \caption{Average total power consumption vs. different delay threshold.}
	\label{PowerDreqgraph}
\end{figure}

\begin{figure}[ht]
	\centering
    \setlength{\abovecaptionskip}{-0.1cm}   
	\includegraphics[scale=0.37]{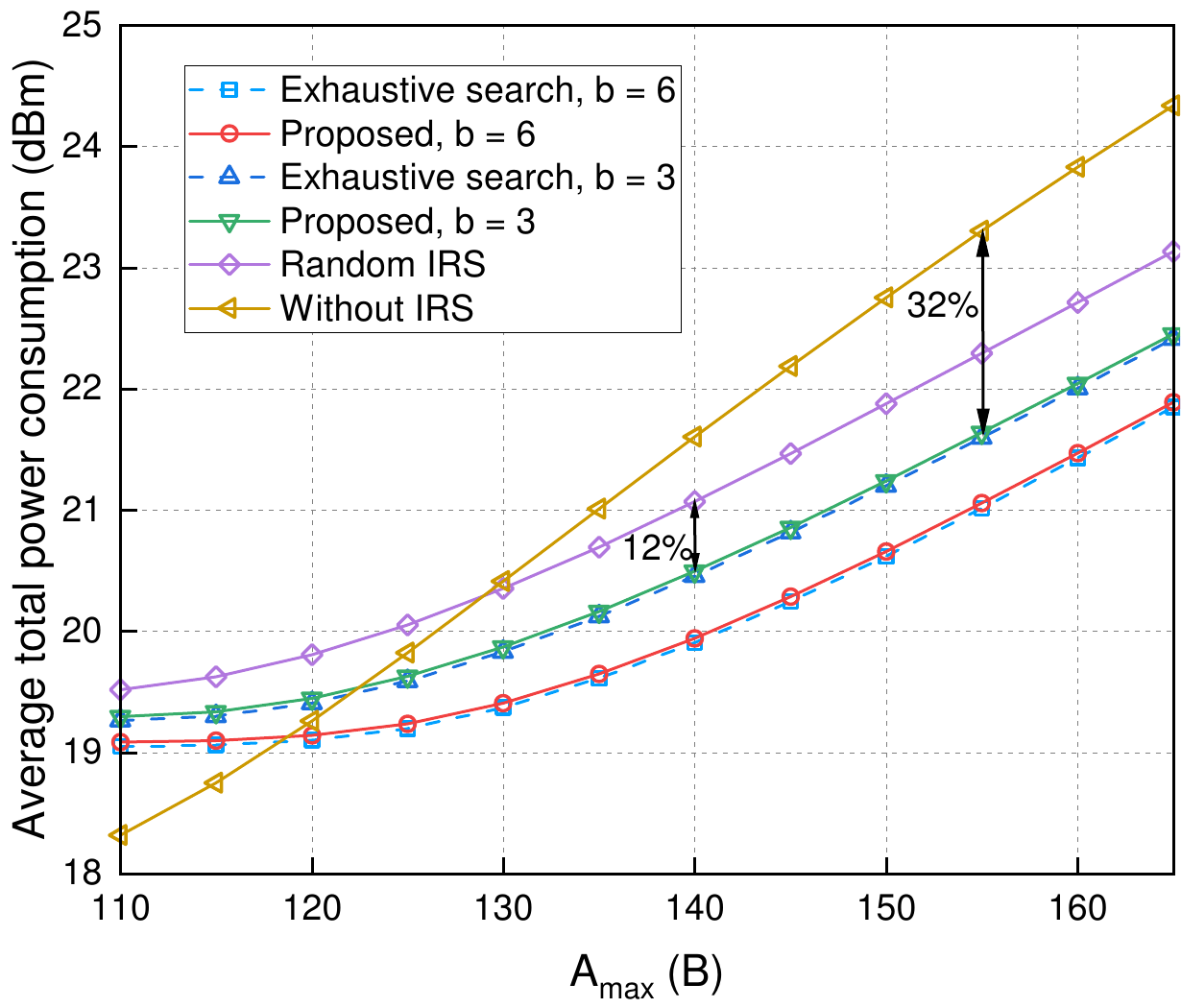}
    \caption{Average total power consumption vs. $A_{max}$.}
	\label{PowerAgraph}
\end{figure}

We initially investigate the convergence of the proposed algorithm with different number of the IoVT devices, as illustrated in Fig. \ref{convergencegraph}. Note that we focus on the convergence performance of the first time slot. As anticipated, the algorithm we proposed convergence quickly across diverse scenarios, thereby substantiating the efficiency of the proposed algorithm.

Fig. \ref{PowerBlockVgraph} illustrates the variations in average total power consumption and the average length of the virtual queue in response to changes in the control parameter $V$. It is evident that by increasing $V$, the average length of the virtual queue increases, while the average overall power consumption decreases. This phenomenon arises from the fact that as $V$ escalates, the proposed algorithm exhibits a heightened preference for the second term of the objective function (\ref{objection function 3}), consequently leading to reduced power consumption and an extended queue length, which shows the inherent trade-off inherent in the drift-plus-penalty method. Consequently, to minimize the average total power consumption, there exists a motivation to set $V$ as large as conceivable. However, once $V$ reaches a certain threshold, further increment yields negligible reductions in total power consumption.

In Fig. \ref{PowerKgraph}, we present the average total power consumption when employing the proposed algorithm and the benchmark algorithms under various number of IoVT devices.
Note that the scenario of $b=6$ can be interpreted as involving continuous phase shifts. Unless explicitly mentioned, all references to the performance of the proposed algorithm below pertain to the scenario where $b=3$.
As expected, the proposed algorithm demonstrates significant performance improvements when compared to the benchmark algorithms. Furthermore, the proposed algorithm exhibits near-optimal performance in comparison to the \textbf{exhaustive search} algorithm.
For the instances where $k = 12$ and $k = 16$, our proposed algorithm achieves performance gains of $15\%$ and $38\%$ respectively, when compared to the \textbf{random IRS} algorithm and the \textbf{without IRS} algorithm. Additionally, it is apparent that as $k$ decreases, the performance advantages of our proposed algorithm over the benchmark algorithm gradually diminish. This phenomenon is particularly notable when $k$ is sufficiently small, and the power consumption of the \textbf{without IRS} algorithm becomes minimal. This trend arises from the challenge of aligning the benefits obtained from serving a limited number of IoVT devices with IRS assistance against the power consumption resulting from the IRS itself.

Fig. \ref{PowerMgraph} and Fig. \ref{PowerNgraph} illustrate the average total power consumption under varying number of IRSs and the number of reflecting elements each IRS. It is evident that, in the majority of scenarios, our proposed algorithm yields significant performance improvements compared to the benchmark algorithm. When $M = 3$ and $M = 5$, the proposed algorithm achieves performance enhancements of $13\%$ and $19\%$ respectively, in comparison to the \textbf{random IRS} algorithm and the \textbf{without IRS} algorithm. Similarly, for $N = 20$ and $N = 40$, the proposed algorithm achieves gains of $16\%$ and $19\%$ respectively over the \textbf{random IRS} algorithm and the \textbf{without IRS} algorithm.
Additionally, it is notable that the achieved performance gains exhibit an initial increase followed by a subsequent decrease as $M$ and $N$ increase. This phenomenon can be explained as the impact of channel gains introduced by increasing $M$ and $N$ diminishes, while the influence of IRS power consumption gradually becomes dominant.

Distinct delay thresholds impose diverse performance requirements on IoVT devices and consequently yield varied impacts on the system. Fig. \ref{DavTgraph} illustrates the variations in average delay for IoVT device under different delay thresholds. Over time, the average delay for IoVT device will eventually stabilize. Fig. \ref{PowerDreqgraph} illustrates the average total power consumption under different delay thresholds. As anticipated, our proposed algorithm exhibits noticeable performance enhancements compared to the benchmark algorithm.
When $d^{th}_k = 80$ ms and $d^{th}_k = 130$ ms, $\forall k \in \mathcal{K}$, the proposed algorithm achieves performance gains of $12\%$ and $18\%$ over the \textbf{random IRS} algorithm and the \textbf{without IRS} algorithm, respectively. Moreover, as the delay threshold gradually decreases, the attainable performance gains also diminish progressively. This phenomenon occurs because when there is less emphasis on low-latency requirements for video data transmission, the impact of IRS on channel quality becomes less conspicuous in comparison to the influence of its own power consumption.

In Fig. \ref{PowerAgraph}, the average total power consumption for different values of $A_{k,max}^t$, $k \in \mathcal{K}$, $t \in \mathcal{T}$, is illustrated. Our proposed algorithm exhibits noticeable performance improvements in comparison to the benchmark algorithm. Compared to the \textbf{random IRS} algorithm and the \textbf{without IRS} algorithm, the proposed algorithm achieves performance gains of $12\%$ and $32\%$ when $A_{k,max}^t = 140$ B and $A_{k,max}^t = 155$ B, $k \in \mathcal{K}$, $t \in \mathcal{T}$, respectively. Moreover, it is observable that as $A_{k,max}^t$ increases, the influence of IRS's own power consumption becomes proportionally smaller compared to the performance enhancement it offers to device performance. Consequently, the performance gains achieved by the proposed algorithm also become increasingly substantial.

\section{Conclusion and Future Work}

In this paper, we propose a multi-IRS assisted IoVT system to achieve low power consumption as well as satisfy delay requirements. To take full advantage of IRS, a long-term total power consumption problem with delay constrained is formulated by jointly optimizing the PBF of IRS and uplink power control at IoVT devices. To solve the problem, we employ Lyapunov optimization to decompose the long-term optimization problem into individual time slots while still meeting the delay constraints. For the non-convex optimization problem within each time slot, an alternative optimization algorithm employing optimal solution seeking and FP is proposed. Finally, extensive simulation results illustrate that our proposed algorithm significantly reduces the long-term total power consumption compared to benchmark algorithms. We have deployed the IRS for the first time in an IoVT system, investigating the potential of IRS to enhance system performance. By deploying an appropriate number of IRS within the IoVT system and utilizing the algorithm we proposed, a significant performance improvement in the IoVT system can be achieved.

It should be emphasized that our research on multi-IRS assisted IoVT systems is still in the preliminary stage, requiring further investigation and exploration.
Firstly, the deployment location of the IRS is a crucial factor affecting system performance. How to deploy IRSs at appropriate locations in IoVT systems to minimize long-term power consumption while satisfying delay requirements is an open topic that needs careful study.
In addition, in practical multi-IRS assisted IoVT systems, perfect channel state information (CSI) cannot be obtained, and how to design joint power control and PBF algorithms with imperfect CSI requires careful investigation.
To facilitate practical deployment further, designing an IRS reflection matrix on BS with imperfect CSI while controlling transmit power at IoVT devices without CSI knowledge presents a significant endeavor.

\bibliographystyle{IEEEtran}
\bibliography{IEEEabrv, Power_Optimization_in_Multi-IRS_Aided_Delay-Constrained_IoVT_Systems}
\end{document}